\documentclass[preprint,showpacs,superscriptaddress,preprintnumbers,amsmath,
amssymb]{revtex4}
\usepackage{graphicx}
\usepackage{subfigure}
\usepackage{dcolumn}
\usepackage{bm}
\usepackage{units}
\begin{document}

\setcounter{page}{1}
\preprint{APS/123-QED}

\title{Study of $K^-$ absorption at rest in nuclei followed by
 $p\Lambda $ emission\footnote{to appear in Phys. Rev. C}}
\author{Grishma Mehta Pandejee}
\email{grishmamehta1982@yahoo.com}
\author{N. J. Upadhyay}
\email{njupadhyay@gmail.com} 
\author{B. K. Jain}
\email{brajeshk@gmail.com}
\altaffiliation[Present Address: ]{UM-DAE, Centre for Excellence in 
Basic Sciences, University of Mumbai, Vidyanagari, Mumbai - 400 098, 
India.}
\affiliation{Department of Physics, University of Mumbai, Vidyanagari,
Mumbai - 400 098, India.}
\date{\today}
\begin{abstract}
$p\,\Lambda$ emission in coincidence following  $K^-$ absorption 
at rest in nuclei is studied using quantum mechanical scattering 
theory and nuclear wave functions. $K^-$ absorption  is assumed  
to occur on two protons in the nucleus. In the formalism, 
emphasis is put on the study of the  final state interaction 
(FSI) effects of $p$ and $\Lambda$ with the recoiling nucleus. 
We include elastic scattering and single nucleon knock-out (KO) 
channels in the FSI. Calculations are presented for the $^{12}$C
nucleus, using shell model wave functions, and without any extra  
mass modification of the $K^-\,pp$ system in the nucleus. 
Calculated results are presented for the angular correlation 
distribution between $p$ and $\Lambda$, their invariant mass 
distribution and the momentum spectra of $p$ and $\Lambda$. 
These results are compared with the corresponding experimental 
measurements \cite{agnello}. With only elastic scattering FSI 
included, the angular correlation distribution and the momentum 
spectra are found to be in good accord with the corresponding 
measurements. With full FSI the calculated $p\,\Lambda$ 
invariant mass distribution is found to have two peaks, one 
corresponding to the elastic scattering FSI and another to single 
nucleon KO FSI. The KO peak agrees fully, in position and shape, with 
the peak observed in Ref. \cite{agnello}. The peak corresponding to 
elastic scattering FSI does not seem to exist in the measured 
distribution. Considering that such a two peak structure is always 
seen in the inclusive ($p$, $p^\prime $) and ($e$, $e^\prime $) 
reactions in nuclei at intermediate energies, absence of the elastic 
scattering peak in the $p\,\Lambda$ reaction is intriguing. 
\end{abstract}


\maketitle

\section{Introduction}
The $K^-\,p$ interaction is attractive in s-wave and isospin, T = 0 
state. Because of this  there is much interest in its study and 
the study of its implications in the possible existence of the  
$K^-$nucleus quasi-bound states in nuclei. Experimentally, the 
existence of such bound states is indicated in the FINUDA 
measurements \cite{agnello} of the stopped $K^-$ absorption on 
Li, C and other target nuclei. These experiments using the FINUDA 
spectrometer installed at the DA$\Phi$NE collider detect a 
$\Lambda$ hyperon and a proton pair in coincidence following $K^-$ 
absorption at rest on several nuclei. The emitted $\Lambda$-$p$ pair 
is found to emerge, predominantly back to back in all target nuclei, 
and have their invariant mass distributions peaking significantly 
below the sum of a kaon and two proton mass in free state (2.370 GeV). 
If it is assumed that the $\Lambda$-$p$ pair is emitted from a 
``$K^-\,pp$" system in the nucleus, this mass shift implies a 
bound $K^-\,pp$ system in nuclei with the binding energy above 100 
MeV. In a more elaborate second run of these experiments carried out 
recently \cite{panic08} it is further reported that these mass 
shifts occur only for the $K^-\,pp$ module, and not for the $K^-\,np$ 
cluster. The absorption on an $n\,p$ pair gives $\Lambda$-$n$ and 
$\Sigma^-$-$p$ pairs in the final state.

Recent analysis of the old DISTO data from the Saturne accelerator 
on $p\,p\,\rightarrow\,p\,\Lambda\,K^+$ reaction too suggests the 
existence of a $K^-\,pp$ cluster with the binding energy around 
100 MeV \cite{disto}.
 
Theoretically, following the extraordinary success of the SU(3) 
chiral perturbation theories in describing the $\pi$-$N$ and 
$K^+$-$N$ systems, the $K^-$-$p$ system has  also been studied under 
this framework, though, unlike pion and $K^+$ cases, the basic 
interaction here is relatively strong. This, however, is 
incorporated in these studies by including  terms up to order 
$q^2$ in the chiral Lagrangian expansion. Then, in combination with 
non-perturbative coupled-channel techniques this framework has been 
found quite appropriate for the study of antikaon-nucleon 
interaction in the literature. It was first developed in Ref. 
\cite{weise1}, and subsequently expanded in Ref. \cite{oolk}. 
Various channels involved for S = $-$1 meson-baryon scattering are 
$\pi^+\,\Sigma^-$, $\pi^0\,\Sigma^0$, $\pi^-\,\Sigma^+$, 
$\pi^0\,\Lambda$, $K^-\,p$, $K^0\,n$. With the proper choice of 
parameters entering in these calculations, all available low-energy 
scattering data in these channels are reproduced well. The $K^-\,p$ 
scattering amplitude resulting from these calculations have a two 
pole structure between $\Sigma\,\pi$ and $\bar{K}\,N$ thresholds. 
The pole which is located close to the real axis couples strongly 
to the $\bar{K}\,N$ channel, while the one coupling strongly to 
$\pi\,\Sigma$ channel lies away from the real axis. Empirically, 
only available information pertaining to $K^-\,p$ scattering below 
threshold is the $\Sigma\,\pi$ invariant mass distribution. This 
mass spectrum has its maximum at 1405 MeV, and has a width of about 
50 MeV. In the Particle Data Group table \cite{pdg} this is 
identified as T = 0, spin half, S = $-$1  $\Lambda(1405)$ $K^-\,p$ 
bound state with a binding energy of 27 MeV. However, it is noticed 
in the literature that there is some subtlety involved in assigning 
a mass to the $\Lambda(1405)$. The observed $\Sigma\,\pi$ spectrum 
has T = 0. Hence, in principle, it can have a generic s-wave T = 0 
source. This means that one needs to fit a superposition of the 
contributions from both, the $\bar{K}\,N$ and $\Sigma\,\pi $ poles 
mentioned above, to reproduce the observed $\Sigma\,\pi $ spectrum 
and assign a mass to $\Lambda(1405)$. Within this scenario it turns 
out that both the poles contribute roughly in equal measure to 
reproduce the measured $\Sigma\,\pi$ spectrum, with a tendency 
towards higher $\bar{K}\,N$ share. A more thorough investigation of 
this issue has been carried out recently in Ref. \cite{weise2}. They 
conclude that for the study of $\bar{K}\,N$ 
scattering, amongst various channels involved in the coupled channel 
calculations, the $\bar{K}\,N$ and $\Sigma\,\pi$ channels dominate
and they couple strongly. They also conclude that the mass of the 
$\Lambda(1405)$ state is in fact 1420 MeV, and not 1405 MeV, thus 
making it only 12 MeV below $\bar{K}\,N$ threshold.

Recent measurements on the $p\,p\,\rightarrow\,p\,K^+\,Y^0$ reaction 
at COSY \cite{zychor}, however, seem to present experimental evidence 
which do not support the above two pole model for the $\Lambda(1405)$. 
The shape and position of the $\Lambda(1405)$ distribution in these 
measurements is reconstructed cleanly in the $\Sigma^0\,\pi^0$ channel 
using invariant- and missing-mass techniques. The mass of the 
$\Lambda(1405)$ is found to be $\sim$ 1400 MeV and width $\sim$ 
60 MeV.  
   
Theoretical search for the antikaon-nucleus bound state has been 
carried out in the literature following the variational approach 
\cite{yama1,dote} and the Faddeev method \cite{shev} for $K^-$ plus 2-3 
nucleons and heavier nuclei. All these calculations need, as input, 
realistic choice for the $NN$ and the $K^-$-$N$ potentials. For the 
nucleon-nucleon potential, following extensive work over the years 
on this subject, it is always possible to make a correct choice. 
For the $K^-$-$N$ potential, however, the situation is  uncertain. 
Some calculations generated a pseudo-potential for it by reproducing 
the $K^-$-$p$ bound state of 27 MeV binding energy, while others used 
a leading order chiral interaction. They all found $K^-\,pp$ bound 
states with about 50 MeV or more binding energies. Latest calculation 
in Ref. \cite{dote}, which, following the two pole model, uses the 
$K^-$-$N$ effective potential corresponding to 1420 MeV mass of 
$\Lambda(1405)$, finds a $K^-\,pp$ system bound by around 19 MeV 
only. This state has a width between 40 and 70 MeV. This suggests 
that the $K^-\,pp$ module might not be sufficiently bound to produce 
any experimentally observable signal corresponding to 
the antikaon-nucleus mesic bound state.

The situation is further confused  by the  suggestion in Ref. 
\cite{magas}, that the down-shift observed in the invariant mass of 
the $p$ and $\Lambda$ in the FINUDA experiment could be the result 
of the final state interaction of these particles with the recoiling 
nucleus. This seems quite plausible because, due to Q-value of the 
$K^-\,+\,p\,p\,\rightarrow\,\Lambda\,p$ reaction being around 
317 MeV, the kinetic energies of outgoing $p$ and $\Lambda$ are 
160 MeV or so. At these energies, it is well known that in the 
nucleon-nucleus scattering the reactive cross section mainly 
consists of the single nucleon knock-out channel \cite{wall}. Thus, 
the knock-out of one nucleon in the nucleus by the out going $p$ or 
$\Lambda$ can shift their energies considerably. The calculations in 
Ref. \cite{magas} indeed reproduce the observed mass shifts in the 
FINUDA experiments. However, as mentioned earlier, the similar 
effects not observed in Run (2) of FINUDA in $K^-$-$p\,n$ absorption 
can not be reconciled with this explanation \cite{panic08}. 

Therefore, the situation on the ($K^-$, $p\,\Lambda$) reaction in the 
nucleus  seems very confusing. It calls for more studies on the 
description of the reaction dynamics, as well as the  $K^-$-nuclear 
binding. 

In the present paper we reexamine the hypothesis of Magas {\it et al.} 
\cite{magas} of the origin of the observed $\Lambda$-$p$ peak in the 
FINUDA measurements to the single nucleon knock-out events in the 
final state. The calculations reported in Ref. \cite{magas} are the 
computer simulations of the internuclear cascade model for the 
nucleon-nucleus scattering. This approach describes the sequence 
of nucleon-nucleon collisions of the outgoing nucleon while passing 
through the residual nucleus in the final state in the framework of 
classical physics. The trajectory of each nucleon is followed. 
After a mean free path a $N$-$N$ collision takes place and its 
results are computed by Monte Carlo or some similar method. Apart 
from the Pauli principle there are no quantum mechanical effects in 
this approach of describing the FSI. The nucleus too is described by 
the Fermi gas model, thus being totally devoid of any nuclear 
structure effect. In view of the crucial role played by the FSI in 
interpreting the FINUDA $\Lambda$-$p$ measurements for $K^-$-nuclear 
bound states, it is absolutely necessary that the FSI in this 
reaction is described using quantum mechanical  
scattering theory and  nuclear wave functions. This is 
the purpose of the present paper.                      

The paper is organized as follows. In Section 2, first, based on 
physical reasoning, we present an overall description of the  
$(K^-\,,\,\Lambda\,p)$ reaction following $K^-$ absorption in the 
nucleus along with an  appropriate final state interaction. This 
is followed by the  formalism utilized for the evaluation of the cross 
section. Calculated results along with a discussion around them are 
presented for $^{12}$C target nucleus. This is followed by the 
conclusions. 

To remain specific in our discussion and presentation, in the 
following we  consider $^{12}$C target nucleus all along.

\section{General Discussion}
\subsection{$K^-$ absorption}    
The $K^-$ meson after being captured in a high atomic orbit reaches 
the $3d$ orbit through electromagnetic transitions. From there onward 
it comes under the influence of the strong nuclear interaction and 
gets absorbed. The X-ray transition width for the $3d$ $\rightarrow$ 
$2p$ transition and the nuclear capture from the $3d$ orbit are 
reported in \cite{back} to be 0.0749 eV and 0.98 $\pm$ 0.19 eV 
respectively. This gives the relative population of kaons in $2p$ and 
$3d$ orbits, using
\begin{equation}
\frac{P(2p)}{P(3d)}\,=\,\frac{\Gamma _X(3d\,\rightarrow\,2p)}
{\Gamma _X(3d\,\rightarrow\,2p)+\Gamma _a(3d)}
\end{equation}
around 7$\%$. Such a small population of kaons in the $2p$ orbit 
also makes nuclear capture of kaons from the $1s$ orbit as 
insignificant. We, thus, consider in our calculations the capture of 
the kaon from both the $3d$ and $2p$ atomic orbits. The $K^-$ 
absorption yield for  the $p\Lambda $ branch is written as the 
weighted average yield from these orbits as
\begin{equation}
\omega\,(p\Lambda)\,=\,\omega_{3d}(p\Lambda)\,+\,0.07\,
\omega_{2p}(p\Lambda).
\end{equation}  
However, as we will see later (Fig. 1), due to larger centrifugal 
barrier, the overlap of the $3d$ orbit with the nuclear wave functions 
is about 2 orders of magnitude smaller than that of the $2p$ orbit. 
Furthermore, as the calculation of the absorption yield, 
$\omega$ involves the square of this overlap, despite the factor of 
0.07 in the above for the $2p$ orbit, the contribution from the $3d$ 
orbit to the $K^-$ nuclear capture remains about two orders of 
magnitude smaller than that from the $2p$ orbit. Therefore, in the 
following we present calculations considering the capture from the 
$2p$ orbit only.    
            
In the absorption process, the $\Lambda$ hyperons in the final state 
are produced either in the quasi-free process $K^-\,N\,\rightarrow\,
\Lambda\,\pi$ or the two-nucleon absorption process $K^-\,N\,N\,
\rightarrow\,\Lambda\,N$. For the  measurements which involve 
$p$-$\Lambda$ detection in coincidence, obviously only the two-nucleon 
absorption contributes. The absorption on more than two nucleons is 
expected to be  weak  because the probability of finding three 
or more nucleons together in the nucleus is small. The Q-value of the 
$K^-\,N\,N\,\rightarrow\,\Lambda\,N$ process (ignoring nuclear 
binding of the absorbing protons) is 317 MeV. This  energy is mainly 
shared by the emerging nucleon and the $\Lambda$ hyperon. Furthermore, 
since the Fermi momentum of the nucleons in the nucleus is not large, 
the outgoing nucleon and the $\Lambda$ hyperon following $K^-$-absorption 
at rest emerge back to back and have their  momenta centred around 
570 MeV/c. Of course, due to Fermi motion, in an actual situation 
the back to back correlation is smeared into two narrow cones, 
and momenta of the nucleon and the $\Lambda$ are spread around 570 MeV/c 
by the Fermi motion. Additionally, the target nucleus, after 
absorption, is left into a two hole shell model state.

Structurally, two aspects of nucleon motion in the nucleus appear 
in the $(K^-\,,\,p\,\Lambda)$ absorption process in the nucleus.   
Because of the predominantly back to back emission of $p$ and 
$\Lambda$ hyperon and their momenta being centred around 570 MeV/c, 
the absorbing pair of nucleons in the nucleus needs to be as close 
as around 0.2 fm to each other at the time of $K^-$ absorption. 
Since the $N$-$N$ potential at these distances is very strong, 
the relative wave function of these protons has strong short range 
correlations. These correlations in the nucleus, however, heals 
very fast \cite{brown} and the wave function goes over to the 
shell model mean field wave function. The centre of mass of these 
two protons, however, has no such constraints on it, hence, it 
always moves in the most probable trajectory given by the nuclear 
mean field. The appropriate 2-proton wave function in the nucleus 
for the kaon absorption, therefore, has the form, 
\begin{equation}
\Psi_{pp}(\vec{r_1}\,,\,\vec{r_2})\,=\,\psi_p(\vec{r_1})\,
\psi_p(\vec{r_2})\,f(r),
\end{equation}
where $\psi$'s are the shell model wave functions and $f(r)$ is a 
Jastrow-type correlation function \cite{jastrow}. 

The outgoing $p$ and $\Lambda$ will also be correlated similarly by 
a correlation function, say $f^\prime(r)$. The healed $p$-$\Lambda$ 
wave function here, however, will be a phase shifted wave function. 

Consequent to the above completely two different space scales involved, 
the absorption probability for $K^-$ in the nucleus for the 
$p$-$\Lambda$ branch (for absorption on a $p\,p$ pair, say) factors 
into two parts (shown in the next section),
\begin{equation}
\omega_{abs}(\vec{p_p}\,,\,\vec{p_{\Lambda}})\,=\,g_{abs}(q)\,
G(Q),
\end{equation}
where $\vec{q}\,=\,(m_p\,\vec{p_\Lambda}\,-\,m_\Lambda\,\vec{p_p})
/(m_p\,+\,m_\Lambda)$ and $\vec{Q}\,=\,\vec{p_p}\,+\,\vec{p_\Lambda}$ 
are the centre-of-mass and the total momenta of the $p$ and $\Lambda$ 
respectively. $g$ and $G$ are respectively the absorption strength 
for $K^-\,p\,p\,\rightarrow\,p\,\Lambda$ process in their centre of 
mass and the momentum probability distribution of nuclear wave 
functions corresponding to the total momentum, $Q$. Because of the 
back-to-back emission of $p$ and $\Lambda$, obviously the magnitude 
of $q$ is very large and that of $Q$ is small.  Due to these vastly 
different momentum scales for $q$ and $Q$, over most of the variables 
measured in the $(K^-\,,\,p\,\Lambda)$ reaction, while $G(Q)$ can go 
through a large variation, the factor $g(q)$ does not change much.

\subsection{Final State Interaction}
\begin{figure}[ht]
\begin{center}
\includegraphics[width=16cm,height=7cm]{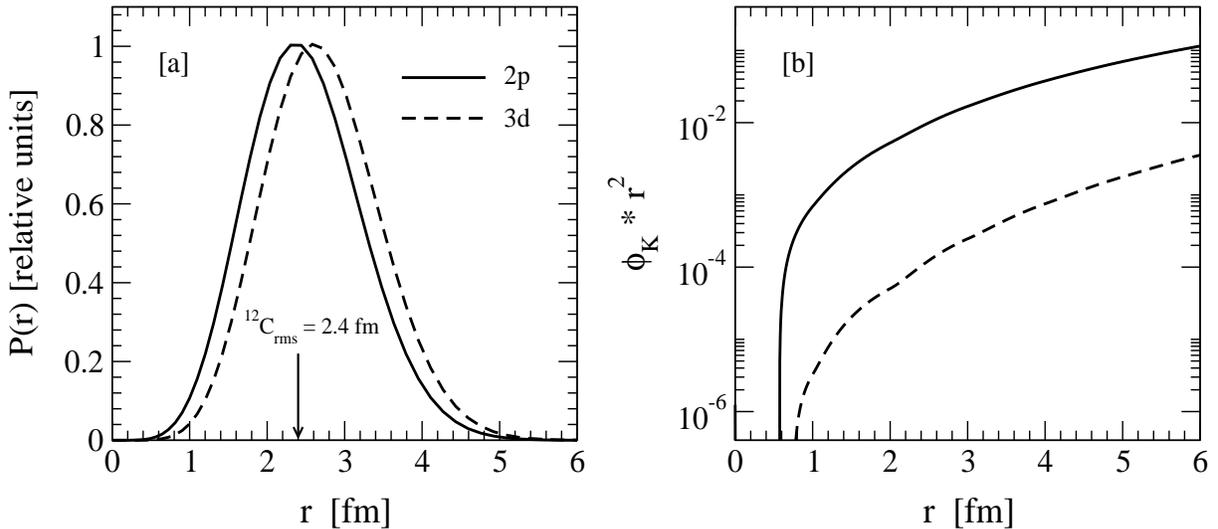}
\end{center}
\caption {(a) Overlap function, $P(r)$ showing the localization of the 
$K^-$ absorption for the $2p$ and $3d$ orbits. The vertical line 
shows the $rms$ radius of the $^{12}$C nucleus. (b) Distribution of the 
$2p$ and the $3d$ $K^-$ wave functions.}
\end{figure}
Before we talk about the final state interaction (FSI), let us 
mention that the $K^-$ absorption in the nucleus occurs on its 
surface. Quantitatively, this region is determined by the overlap 
of the $K^-$ $2p$ and $3d$ atomic orbits with the spatial distribution 
of the absorbing protons. Fig. 1a shows these overlaps, where  
\begin{equation}
P(r)\,=\,r^2\,\phi_K(r)\,\psi_{p_1}(r)\,\psi_{p_2}(r).
\end{equation}       
For the $^{12}$C nucleus we have taken the two protons moving in 
the $1p$ shell model orbital. Their radial distribution is described 
by the harmonic oscillator wave function. The  kaonic atomic orbits 
are given by the hydrogenic wave function. To show their relative 
localization clearly, in Fig. 1a we plot $P(r)$'s  for the $2p$ and 
the $3d$ $K^-$ orbits, which have different magnitudes, on the same 
scale with arbitrary units. We see that both the overlap functions 
peak around r = 2.5 fm, with $3d$ overlap function about half a fermi 
ahead. The $^{12}$C rms radius is known to be 2.4 fm. The magnitude 
of the $3d$ overlap function, we note is two orders of magnitude 
smaller than that of the $2p$ orbit. This happens because the $3d$ 
wave function rises slower than the $2p$ $K^-$ wave function in the 
region of the overlap, as we see in  Fig. 1b. 

Above localization of the $P(r)$ beyond the $^{12}$C rms radius
coupled with the back to back emission of $p$ and 
$\Lambda$ following absorption, suggests that only one of the 
two emitted particles, $p$ and $\Lambda$, goes into the nucleus 
at a time. The other particle moves outward. Hence, the  
 FSI with the recoiling nucleus is mainly suffered  by only   
one particle, $p$ or $\Lambda$ at a time. The channels which 
dominate in contribution to this interaction  are the elastic 
channel and the single nucleon knock-out (KO) channel. The latter 
is known to constitute about 80$\%$ of the total reactive cross 
section in the proton-nucleus inelastic scattering in nuclei around 
160 MeV proton energy \cite{wall}, which is the relevant proton 
energy in the present study. Out of these, the effect of the elastic 
channel at intermediate energies is mostly absorptive, while that of 
the KO channel is dispersive as well as absorptive. In our 
calculations, we include both the channels. The inclusive probability 
for a process like 
\begin{equation}
K^-\,+\,A\,\rightarrow\,p\,+\,\Lambda\,+\,X,
\end{equation} 
is therefore written as
\begin{equation}
d\omega\,=\,d\omega_{elas}\,+\,d\omega_{KO}.
\end{equation}
The sum of two terms in the above is incoherent because in principle 
(by making exclusive measurements) we can distinguish between 
elastic scattering and single scattering with one target nucleon 
being knocked out.

\section{Formalism and Results}  
\subsection{Elastic}
\begin{figure}[ht]
\begin{center}
\subfigure[]{\includegraphics[width=6.0cm,height=4.5cm]{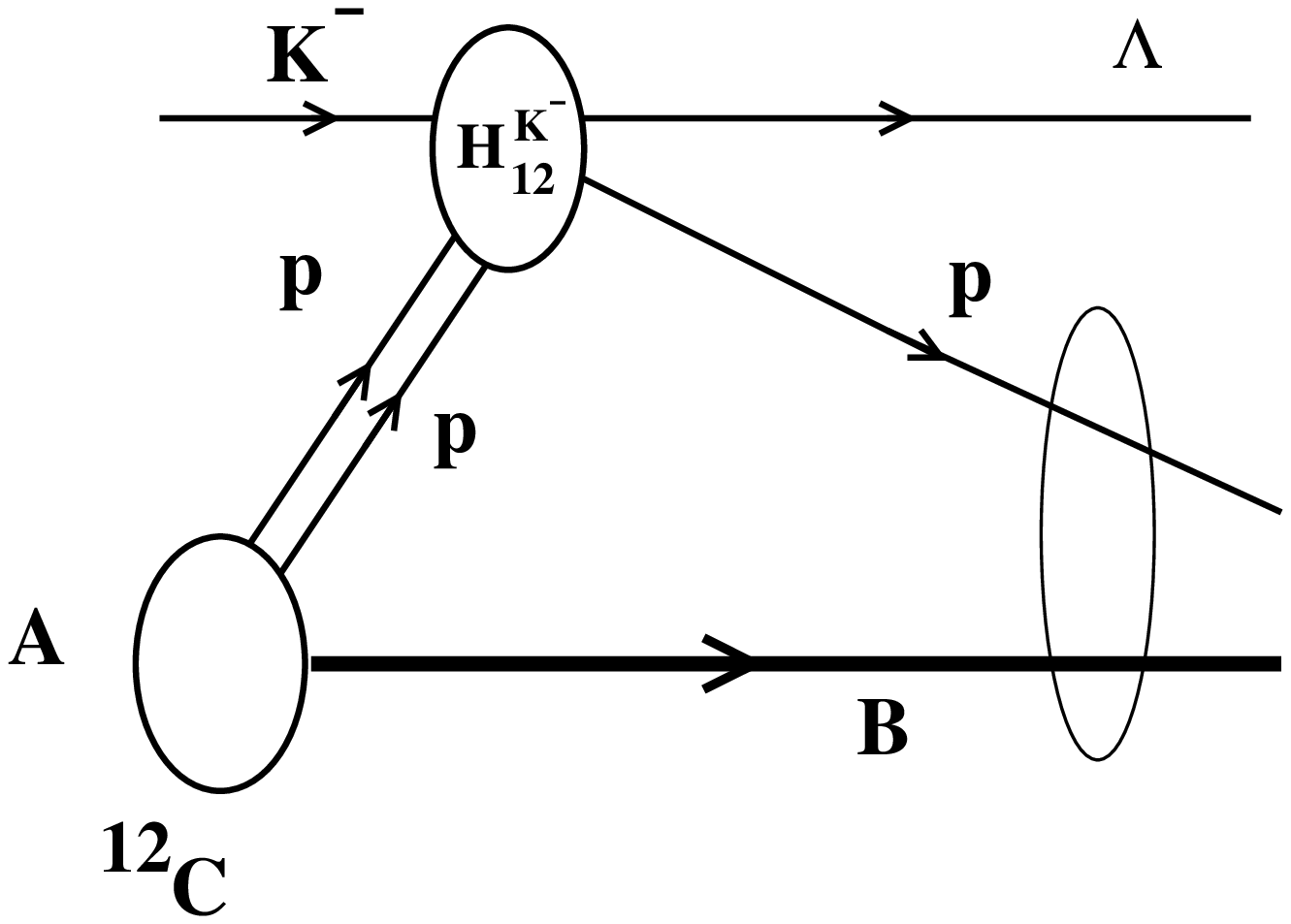}}
\subfigure[]{\includegraphics[width=6.0cm,height=4.5cm]{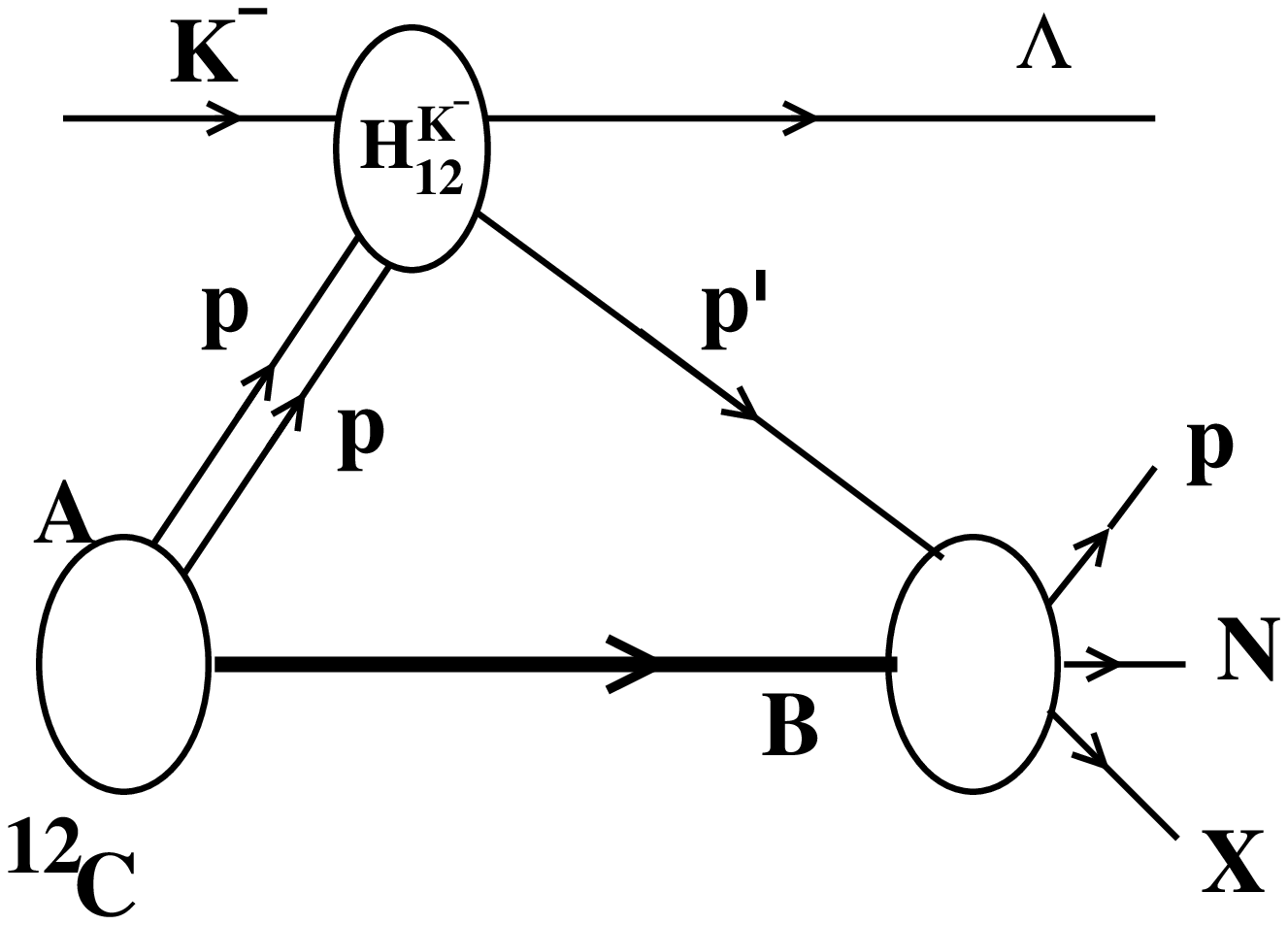}}
\end{center}
\caption {Reaction mechanism of (a) Elastic process and (b) Knock-out 
process.}
\end{figure}
The ``elastic" process for the $A\,(K^-\,,\,p\,\Lambda)\,B$ reaction 
is shown in Fig. 2a. In it the $K^-$ gets absorbed on a pair of 
protons in the target nucleus and produces a $p$-$\Lambda$ pair.  
This pair is detected in coincidence in the final state. No 
measurements are made on the recoiling nucleus, hence the 
measurements are inclusive in that sense. The recoiling nucleus is 
left in a {\it two hole} state centred around the excitation 
corresponding to the summed binding energy ($B_1\,+\,B_2$) of 
nucleons 1 and 2 in the nucleus A. Let us denote different states of 
B around this excitation by $n$. The inclusive absorption 
probability  for  protons in shells ($n_1\,l_1$; $n_2\,l_2$) in the 
nucleus is then given by
\begin{eqnarray}
\nonumber
d\omega_{elas}&=&\frac{1}{(2\pi )^5}\,\,\sum_n\,\,\delta(M_i\,-\,
T_\Lambda\,-\,T_p\,-\,T_B\,-\,E_n^*)\,\,\delta(\vec{P_B}\,+\,
\vec{p_p}\,+\,\vec{p_\Lambda})
\\
&&\hspace{3.5cm}\times\,\,d\vec{p_p}\,\,d\vec{p_\Lambda}\,\,
d\vec{P_B}\,\,\bar{\sum _\sigma}\,|M_{fi}|^2,
\end{eqnarray}
where $M_i\,=\,m_K\,+\,m_p\,-m_\Lambda $ and $E_n^*$ is the 
excitation energy of the state $n$ in $B$. $T_x$ denotes the kinetic 
energy of the particle $x$. Bar on the sum in the above expression 
denotes the average and sum over the spins in the initial and final 
states respectively. The transition matrix element $M_{fi}$ is given 
by 
\begin{equation}
M_{fi}\,=\,[N_{l_1\,l_2}]^{1/2}\,\int d\xi\,\,dx_1\,\,dx_2\,\,
\Psi_{B\,,\,n}^*(\xi)\,\,\chi^{-*}(x_1,x_2)\,\,H_{12}^{K^-}\,\,
\Psi_A(\xi,x_1,x_2)\,\,\phi_K,
\end{equation}
where $N_{l_1\,l_2}$ are the active number of absorbing proton pairs 
in shell ($n_1\,l_1$; $n_2\,l_2$) in the target nucleus, $A$. 
$\Psi_x$ is the nuclear wave function, and $\chi$ is the elastically 
scattered $p$ and $\Lambda$ wave function. $ H_{12}^{K^-}$ is the 
absorption vertex, and it depends only on the proton coordinates, 
$x_1$ and $x_2$ in the nucleus A. $\xi$ represents collectively the 
coordinates of the $(A-2)$ nucleons.
    
To proceed further we note that since the excited states `$n$' in the 
nucleus $B$ are the hole states corresponding to two nucleons, they 
are not likely to have much energy spread. Hence,  in the energy 
delta function in Eq. (8) we replace $E_n^*$ by $|\,B_1\,+\,B_2\,|\,
=\,B_{12}$. With this we obtain     
\begin{equation}
d\omega_{elas}\,=\,[PS]\,\sum_n \bar{\sum _\sigma}\,|M_{fi}|^2,
\end{equation}
with  $[PS]$, 
the phase space factor,  given by 
\begin{equation}
[PS]\,=\,\frac{1}{(2\pi )^5}\,\,\delta(M_i\,-\,T_\Lambda\,-\,T_p\,
-\,T_B\,-\,B_{12})\,\,d\vec{p_\Lambda}\,\,d\vec {p_p},    
\end{equation}
and $\vec{P_B}\,=\,-\,(\vec {p_p}+\vec{p_\Lambda})=-\vec {Q}$.
Sum over `$n$' is now performed using the ``closure relation", 
yielding 
\begin{eqnarray}
\nonumber
&&\bar{\sum _\sigma}\,\sum_n\,|M_{fi}|^2\,\equiv\,
\bar{| M_{fi}|^2}
\\
&&=\,[N_{l_1\,l_2}]\,\,\bar{\sum _\sigma}\,\int d\xi\,\,\left|\,\int 
dx_1\,\,dx_2\,\,\chi^{-*}(x_1,x_2)\,\,H_{12}^{K^-}\,\,
\Psi_A(\xi,x_1,x_2)\,\,\phi_K\,\right|^2.
\end{eqnarray}
Since we are interested only in the inclusive absorption strength 
we take a simple description of the target nucleus, where the 
absorbing protons move in shell model orbitals  $(n_1\,l_1)$ and 
$(n_2\,l_2)$ and the core of $(A-2)$ nucleons is a spectator. With 
this description, above expression reduces to
\begin{equation}
\bar{|M_{fi}|^2}\,=\,[N_{l_1\,l_2}]\,\,\bar{\sum _\sigma}\,\,\left|\,
\int dx_1\,\,dx_2\,\,\chi^{-*}(x_1,x_2)\,\,H_{12}^{K^-}\,\,
\Psi_{n_1l_1m_1;n_2l_2m_2}(x_1,x_2)\,\,\phi _K\,\right|^2,
\end{equation}
where $\Psi_{n_1l_1m_1;n_2l_2m_2}(x_1,x_2)$ is the properly 
anti-symmetrized two proton wave function in the nucleus. In the $LS$ 
representation it is written as 
\begin{eqnarray}
\nonumber
\Psi_{n_1l_1m_1;n_2l_2m_2}(x_1,x_2)\,=\,\sum_{LMS\sigma } 
(l_1\,\,l_2\,\,m_1\,\,m_2\,/\,L\,\,M)\,\,(\nicefrac{1}{2}\,\,
\nicefrac{1}{2}\,\,\sigma_1\,\,\sigma_2\,/\,S\,\sigma)
\\
\times\,\,\phi_{LM}(\vec{r_1},\vec{r_2})\,\,
\chi_{S\sigma}(\vec{s_1},\vec{s_2}).
\end{eqnarray}
For two protons in the same shell antisymmetry requires that 
$L\,+\,S\,=\,even$. Further on, performing sum over $m_1\,,\,m_2\,,\,
\sigma_1$ and $\sigma_2$ contained in 
$\displaystyle\bar{\sum_\sigma}$ in Eq. (13), we get 

\begin{eqnarray}
&&\bar{|M_{fi}|^2}\,=\,N_{l_1\,l_2}\,\,\frac{1}{4}\,\,\sum_{S\sigma}
\frac{1}{(2l_1+1)(2l_2+1)}\,\times
\\
\nonumber
&&\hspace{1cm}\sum_{LM}\,\,\left|\,\int dx_1\,\,dx_2\,\,
\chi^{-*}(x_1,x_2)\,\,H_{12}^{K^-}\,\,\phi _K\,\,\phi_{LM}
(\vec{r_1},\vec{r_2})\,\,\chi_{S\,\sigma}(\vec{s_1},\vec{s_2})
\right|^2.
\end{eqnarray}
The correlation functions $f(r)$ and $f^\prime(r)$ mentioned in 
Eq. (3) and after it are absorbed here in the absorption vertex 
$H_{12}^{K^-}$.

\subsubsection{Evaluation of $\bar {|M_{fi}|^2}$}
To proceed further, let us now utilize the fact that the momentum 
$\vec{q}$ appearing in the absorption vertex $H_{12}^{K^-}$ has 
large magnitude and a short range. Because of this, in the 
expression for $\bar{|M_{fi}|^2}$ we can factorize the expectation 
value of $H_{12}^{K^-}$ from the rest, and write,
\begin{eqnarray}
&&\bar{|M_{fi}|^2}\,=\,\left[\,\frac{1}{4}\,\,\sum _{S\sigma}
\,\,\left|\,\langle p,\Lambda ,\vec{q}\,|\,H_{12}^{K^-}\,|\,
\chi_{S\sigma}(\vec{s_1},\vec{s_2}),ppK^-\,\rangle\,\right|^2\,
\right]\,\,\times
\\
\nonumber
&&\left[\,\frac{N_{l_1\,l_2}}{(2l_1+1)(2l_2+1)}\,\,\sum_{LM}\,
\left|\,\int d\vec{r_1}\,\,d\vec{r_2}\,\,\chi ^{-*}
(\vec{r_1},\vec{r_2})\,\,\phi_K(\vec{r_1})\,\,\phi_{LM}
(\vec{r_1},\vec{r_2})\,\,\delta(\vec{r_1}-\vec {r_2})\,\right|^2
\,\right].
\end{eqnarray}
Two expressions in the square brackets in above can be identified 
with two terms of Eq. (4), i.e.
\begin{equation}
g_{abs}(q)\,=\,\left[\,\frac{1}{4}\,\,\sum_{S\sigma,\sigma_p,
\sigma_\Lambda} \left|\,\langle p,\Lambda,\vec {q}\,|\,
H_{12}^{K^-}\,|\,\chi_{S\sigma}(\vec{s_1},\vec{s_2}),ppK^-\rangle\,
\right|^2\,\right],
\end{equation}
and
\begin{eqnarray}
G(Q)&=&[PS]\,\,N_{l_1\,l_2}\frac {1}{(2l_1+1)(2l_2+1)}\,\times
\\
&&
\nonumber
\sum_{LM}\,\,\left|\,\int d\vec{r_1}\,\,d\vec{r_2}\,\,
\chi^{-*}(\vec{r_1},\vec{r_2})\,\,\phi_K(\vec{r_1})\,\,
\phi_{LM}(\vec{r_1},\vec{r_2})\,\,\delta(\vec{r_1}-\vec{r_2})\,
\right|^2.
\end{eqnarray}

\subsubsection{Distorted waves $\chi's$}
As the energies of the proton and the  lambda following $K^-$ 
absorption is around 160 MeV or so, we describe the scattering of 
these particles by the recoiling nucleus using eikonal approximation. 
The basic assumption in this description is that the propagating 
particle is mainly scattered in the forward direction. Taking 
z-axis parallel to the proton momentum, $\vec{p_p}$, the  proton 
distorted wave, $\chi_{\vec{p_p}}^{-*}$ is written in eikonal 
approximation as 
\begin{equation}
\chi_{\vec{p_p}}^{-*}(\vec {r})\,=\,e^{-i\vec{p_p}\,\cdot\,\vec{r}}
\,D_{\vec{p_p}}(\vec{r}),
\end{equation}
where the distortion function $D$ is given in terms of an optical 
potential, $V_p$ by
\begin{equation}
D_{\vec{p_p}}(\vec{r})\,\equiv D_{\vec{p_p}}(\vec{b}, z)\,=\,exp\,
\left[\,-\,\frac{i}{\hbar\,v_p}\,\,\int_z^\infty V_p(\vec{b},
z^\prime)\,\,dz^\prime\,\right],
\end{equation}
where $\vec{r}\,=\,(\vec{b},z)$.
For writing the $\Lambda$ distorted wave we recall that the $\Lambda$ 
moves opposite to the proton. Therefore, the momentum vector 
$\vec{p_\Lambda}$ is anti-parallel to the chosen z-axis. The 
distortion factor, $D_{\vec{p_\Lambda}}(\vec{r})$ for $\Lambda$ 
therefore becomes \cite{jackson} 
\begin{equation}
D_{\vec{p_{\Lambda}}}(\vec{r})
=exp\,\left[\,-\,\frac{i}{\hbar\,v_\Lambda}\,\,\int_{-\,\infty}^z 
V_\Lambda(\vec{b},z^\prime)\,\,dz^\prime\,\right].
\end{equation}
Combining $D's$ for the proton and the lambda we then get 
\begin{equation}
D_{\vec {p_p}}(\vec {r})\,\,D_{\vec {p_\Lambda}}(\vec {r})\,=\,
exp\,\left[\,-\,\frac{i}{\hbar}\,\left(\,\int_{-\,\infty}^z 
\frac{V_\Lambda}{v_\Lambda}\,\,dz^\prime\,+\,\int_z^\infty 
\frac{V_p}{v_p}\,\,dz^\prime\,\right)\,\right].
\end{equation}
If we make the $``t\rho "$ approximation for $V's$ and assume 
forward scattering for $t$, we get 
\begin{equation}
D_{\vec{p_p}}(\vec{r})\,\,D_{\vec{p_\Lambda}}(\vec{r})\,=\,
exp\,\left[\,\frac{i}{2}\,\left(\,\sigma_T^{\Lambda N}
(i+\beta_{\Lambda N})\,\,\int_{-\,\infty}^z \rho\,\,dz^\prime\,+\,
\sigma_T^{pN}(i+\beta_{pN})\,\,\int_z^\infty \rho\,\,dz^\prime\,
\right)\,\right],
\end{equation}
where $\sigma_T^x$ and $\beta_x$ are respectively the total cross 
section and the ratio of the real to imaginary part of the 
scattering amplitude for the $xN$ system.
 
Now, if we ignore the difference between  the proton and the lambda 
elementary scattering parameters and take them as that for 
the better studied $p\,N$ system at some mean value of the $p$ and 
$\Lambda$ energies, above expression simplifies to 
\begin{eqnarray}
\nonumber
D_{\vec{p_p}}(\vec {r})\,\,D_{\vec{p_\Lambda}}(\vec{r})\,\equiv\, 
D(\vec{r})&=&exp\,\left[\,\frac{i}{2}\,\,\sigma_T^{pN}(i+\beta_{pN})
\,\,\int_{-\,\infty}^\infty \rho(\vec{r^\prime})\,\,dz^\prime\,
\right]
\\
&=&exp\,\left[\,\frac{i}{2}\,\,\sigma_T^{pN}(i+\beta _{pN})\,\,
T(\vec{b})\,\right],
\end{eqnarray}
where $\rho(r)$ is the nuclear density  and $T(\vec{b})$ 
is the total nuclear material seen by the proton and the lambda 
{\it together } at an impact parameter, $\vec{b}$. It is given by
\begin{equation}
T(\vec{b})\,=\,\int_{-\,\infty}^\infty \rho(\vec{r^\prime})\,\,
dz^\prime,
\end{equation}
with $r^\prime\,=\,\sqrt{b^2+{z^\prime}^2}$. Now if we observe 
Eq. (24) for $D(\vec{r})$ a little closely, we realize that this, 
in fact, is the total distortion factor for the passage of a 
proton from  one end of the nucleus to  another. This, thus, 
is the mathematical description for the statement made in an earlier 
Section that, because of the peripheral localization of the $K^-$ 
absorption and  the back-to-back emission of $p$ and $\Lambda$, total 
scattering of the $p$ and the $\Lambda$ can be included in the final 
state by considering the  passage of only one particle ($p$ or 
$\Lambda$) through the whole nucleus.

Furthermore, since the effect of distortion at the energies of $p$ 
and $\Lambda$ in the studies here ($\sim$ 160 MeV) is mainly absorptive, 
and most of the measurements on $p\Lambda$ following $K^-$absorption 
are of inclusive type, it will be reasonably correct to include the 
overall effect of the distortion in $G(Q)$ by multiplying it by an 
attenuation factor, $\eta_A(T_p)$, given by   
\begin{eqnarray}
\nonumber
\eta_A(T_p)&=&\frac{\int d\vec{b}\,\,dz\,\rho_A(\vec{b},z)
\,|\,D(\vec{r})\,|}{\int d\vec{b}\,\,dz\,\rho_A(\vec{b},z)}
\\
&=&\frac{\int d\vec{b}\,\,dz\,\rho_A(\vec{b},z)\,
e^{-\frac{1}{2}(P\sigma_T^{pN}(T_p))\,\,
T(\vec{b})}}{\int d\vec{b}\,\,dz\,\rho_A(\vec{b},z)},
\end{eqnarray}     
and removing the distortion factor from the integral in Eq. (18).
Factor $P$ in above has been introduced before $\sigma_T^{pN}$ to 
include the effect of Pauli-blocking of the nucleons in the nucleus 
after scattering. A nuclear matter  estimate for its value above 
twice the Fermi energy is given by \cite{fermi}
\begin{equation}
P(\epsilon)\,=\,1\,-\,\frac{7}{5\epsilon},
\end{equation}
where $\displaystyle \epsilon\,=\,\frac{T_p}{E_F}$, with $E_F$ 
denoting the Fermi energy.

With above treatment of distortion, Eq. (18) for $G(Q)$ simplifies to 
\begin{equation}
G(Q)\,=\,[PS]\,\,|\eta_A(T_p)|^2\,\,\sum_{LM}|\,F_{LM}^{l_1l_2}(Q)\,
|^2,
\end{equation}
where 
\begin{eqnarray}
\nonumber
\sum_{LM}|\,F_{LM}^{l_1\,l_2}(Q)\,|^2&=&\sum_{LM}\left|\,
\sqrt{\frac{N_{l_1\,l_2}}{(2l_1+1)(2l_2+1)}}\,\,\int d\vec{r_1}\,
\,e^{-i\,\vec{Q}\,\cdot\,\vec{r_1}}\,\,\phi_K(\vec{r_1})\,\,
\phi_{LM}(\vec{r_1},\vec{r_1})\,\right|^2
\\
&=&N_{l_1\,l_2}\,\,\sum_L (l_1\,l_200\,/\,L\,0)^2\,\,|\,
g_{l_1\,l_2}^L(Q)\,|^2,
\end{eqnarray}
with
\begin{equation}
g_{l_1\,l_2}^L(Q)\,=\,\int dr\,\,r^2\,\,j_L(Qr)\,\,R_{n_1\,l_1}(r)
\,\,R_{n_2\,l_2}(r)\,\,\phi_K(r),
\end{equation}
where $R_x$ are the nucleon radial wave functions in the nucleus.
\begin{figure}[ht]
\begin{center}
\includegraphics[width=11cm,height=9cm]{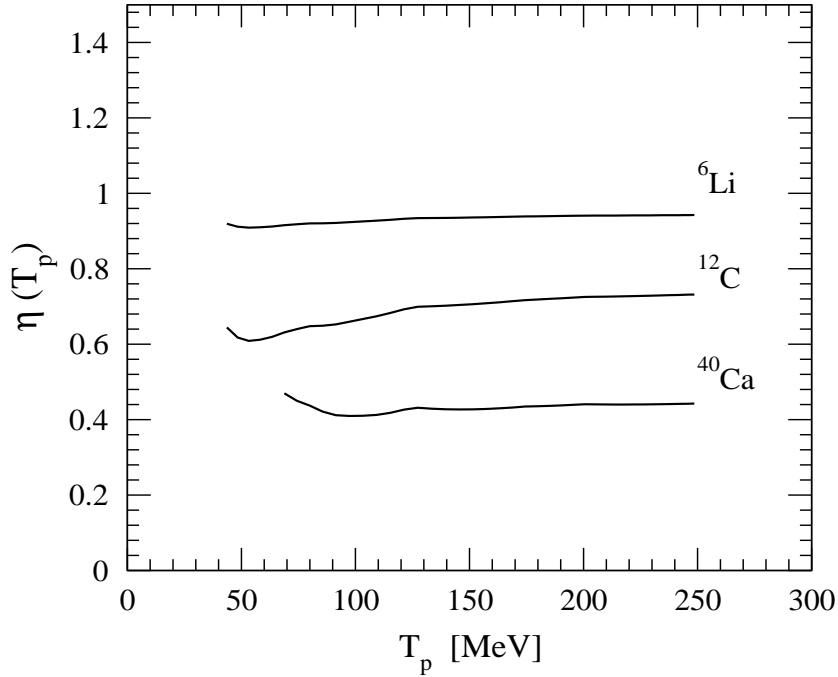}
\end{center}
\caption {Attenuation factor $\eta_A(T_p)$ (Eq. (26)) for $^6$Li, 
$^{12}$C and $^{40}$Ca nuclei.}
\end{figure}

The value of $\eta_A(T_p)$ depends upon the proton energy, $T_p$ 
through $P\sigma_T^{pN}$ and upon the nucleus through the 
thickness function, $T(b)$. To get an idea about the value 
and its variation, in Fig. 3 we plot $\eta_A(T_p)$ in the 
proton kinetic energy range 50-250 MeV for three nuclei, $^6$Li, 
$^{12}$C and $^{40}$Ca. For calculating the Pauli blocking factor, 
$P$, Fermi momentum is taken equal to 200 MeV/c. The nuclear 
densities are taken from Ref. \cite{den}. We find that in the 
energy range (100-200 MeV) of interest in the present calculations 
the attenuation factor approximately remains constant. The values 
of this factor for the three nuclei are around 0.9, 0.7 and 0.4 
respectively.

\subsubsection{Absorption Vertex, $g_{abs}(q)$}
Prescription to describe the absorption vertex, $g_{abs}(q)$
is not clear and also not simple. One thing, which is definite about 
it is that, it involves large momentum transfer, hence, spatially 
it can not be localized over any extended volume. Dynamically, the 
one-nucleon absorption mechanism, $K^-p\rightarrow \pi \Lambda $ is 
understood to involve the $\Lambda (1405)$, which decays in to a pion 
and a hyperon. The range of this vertex is determined by the 
$\Lambda (1405)$ propagator. A study in Ref. \cite{rook} suggests 
that the absorption probability depends upon this range, and in their 
estimate this range could be around 0.75 fm or so. The two nucleon 
mechanism, due to strong attractive $K^-$-$p$ interaction in the T = 0 
state, involves  dynamically a strongly correlated system of $K^-pp$, 
where the $K^-$ is continuously exchanged between two protons. 
Detailed dynamical composition of this system is determined in the 
$\chi $PTs by the non-perturbative coupling amongst various S = $-$1, 
T = 0 channels, $\pi^+\,\Sigma^-$, $\pi^0\,\Sigma^0$, $\pi^-\,
\Sigma^+$, $\pi^0\,\Lambda$, $K^-\,p$, $K^0\,n$. This system 
eventually decays in to $p\Lambda$. Exact mechanism of this decay is 
not immediately obvious. However, in line with the one-nucleon $K^-$ 
absorption, one mechanism could be that at some stage in the $K^-pp$ 
system a $\Lambda (1405)$ is produced, which, as suggested in 
Ref. \cite{dote}, interacts with another proton through an exchange 
of pions or a pair of $\pi$-$K$ and goes over to the $p$-$\Lambda$ final 
state. In Ref. \cite{dote}, using the range parameter 0.2- 1.2 fm for 
the absorption vertex the authors estimate the decay width of 2-8 MeV 
for the $K^-pp$ system, with the maximum width occurring for the 
range around 0.7 fm. Thus, it appears that, even if the details of 
the absorption vertex is not known very clearly, two things are 
clear: (i) the magnitude of the two proton $K^-$ capture depends upon 
the spatial extension of the absorption vertex, and (ii) the probable 
range of the vertex is such that the variation of the capture 
probability with momentum, $q$ could not be very rapid.      
  
The ($K^-$, $p\,\Lambda$) reaction measurements have been done on the 
distributions of the $\Lambda $ (and proton) momentum, invariant 
$p\,\Lambda$ mass, and their angular correlations. In all these 
distributions, as we will discuss in the next Section, each point 
involves a folding of $g(q)$ and $G(Q)$. However, the values of 
$q$ for all the measurements are large (around 500 MeV/c) and do not 
have much spread (only up to 10$\%$). Because of this, the factor 
$g_{abs}(q)$ (Eq. (17)) in the formalism enters in determining the 
absolute magnitude of the ($K^-$, $p\,\Lambda$) process only. The 
shapes of different distributions are determined by the factor $G(Q)$. 
Furthermore, as the available data from the experiments exist only 
in arbitrary units, we have taken $g_{abs}(q)$ as a constant factor, 
denoted by $C$, in our calculations. 

\subsubsection{Results}
Before we present the results let us make some points about the FINUDA 
experiment. The $\Lambda$'s are detected in this experiment by 
reconstructing the invariant mass of the $\Lambda$  decay products, 
$p$ and $\pi ^-$. However, the restriction on the low momentum 
threshold for $\pi ^-$ in the FINUDA spectrometer is such that it cuts 
out the $\Lambda $ hyperons with a momentum lower than 300 MeV/c. 
Therefore, the $\Lambda $ from the quasifree process ($K^-\,N\,
\rightarrow\,\Lambda\,\pi$) in this experiment is hardly observed. 
Above around 400 MeV/c , since the main contribution comes from 
two-nucleon absorptions ($K^-\,``NN"\,\rightarrow\,\Lambda\,N\,,\,
\Sigma^0\,N$), the FINUDA measurements have major contribution from 
this capture process. Furthermore, the measurements are done with the 
$p$ and $\Lambda $ in coincidence, with the cosine of the angle 
between them restricted as $-1\,\leq$ cos $\theta_{p\Lambda}\,\leq\,
-0.8$. The latter constraint is put because the cross section beyond 
these limits is insignificant (see Fig. 4).    
 
For calculations, putting all factors together from the above Section, 
the differential absorption strength, $d\omega$ for $K^-$-absorption 
on two protons in shell model orbitals, ($n_1\,l_1$; $n_2\,l_2$) is 
finally written as 
\begin{equation}
d\omega_{elas}\,=\,[PS]\,C\,|\eta_A(T_p)\,|^2\,N_{l_1\,l_2}
\,\,\sum_L(l_1\,l_2\,0\,0\,/\,L\,0)^2\,\,|\,g_{l_1\,l_2}^L(Q)\,|^2,
\end{equation}             
where all the terms are as defined in above Sections. We present here 
calculated  results using this expression  for the   
$^{12}$C target nucleus with an appropriate phase space factor, $[PS]$. 
The proton wave functions in $^{12}$C are 
generated in an oscillator potential, whose length parameter, $b$ is 
taken equal to 1.67 fm. This parameter fits the elastic electron 
scattering data on the $^{12}$C nucleus \cite{den}. The binding 
energies of protons in $1p$ and $1s$ shells are taken as given by 
the ($p$, $2p$) and ($e$, $e^\prime\,p$) reactions \cite{be}. For 
$^{12}$C they are 15.96 and 34.0 MeV respectively. The $K^-$ 
absorption is considered to occur on proton pairs in $1p$, $1s$ 
and $1s1p$ shells. The value of $N_{l_1\,l_2}$ for these absorptions 
is taken equal to the number of possible proton pairs in these 
shells, as an upper limit. They are in the ratio of 
$(1p)^2:(1s1p):(1s)^2::6:8:1$. 

The phase space factor, $PS$ for calculating the angular correlation 
between the $p$ and $\Lambda $ and their momentum distribution is 
written as 
\begin{equation}
PS\,=\,\frac{4\,\pi\,m_p\,m_{\Lambda}}{(2\pi)^5}\,\,
\frac{|\vec{p_p}|^2|\,\vec{p_\Lambda}|^2}{(m_{\Lambda}\,+\,m_B)\,
|\vec{p_{\Lambda}}|\,+\,|\vec{p_p}|\,m_{\Lambda}\,cos\,
(\theta_{\Lambda\,p})}\,d\,cos\,(\theta_{\Lambda\,p})\,d\,
|\vec{p_{\Lambda}}|.
\end{equation}.

In Fig. 4 we present the calculated angular correlation between 
the $\Lambda$ and the proton along with the FINUDA measurements. 
We find them to agree very well with each other. The steep rise 
in the distribution towards $\theta_{\Lambda p}$ = 180$^{\circ}$ 
is the strong confirmation of the two-proton absorption mechanism. 
Both the results are given in relative units.    
\begin{figure}[ht]
\begin{center}
\includegraphics[width=11cm,height=9cm]{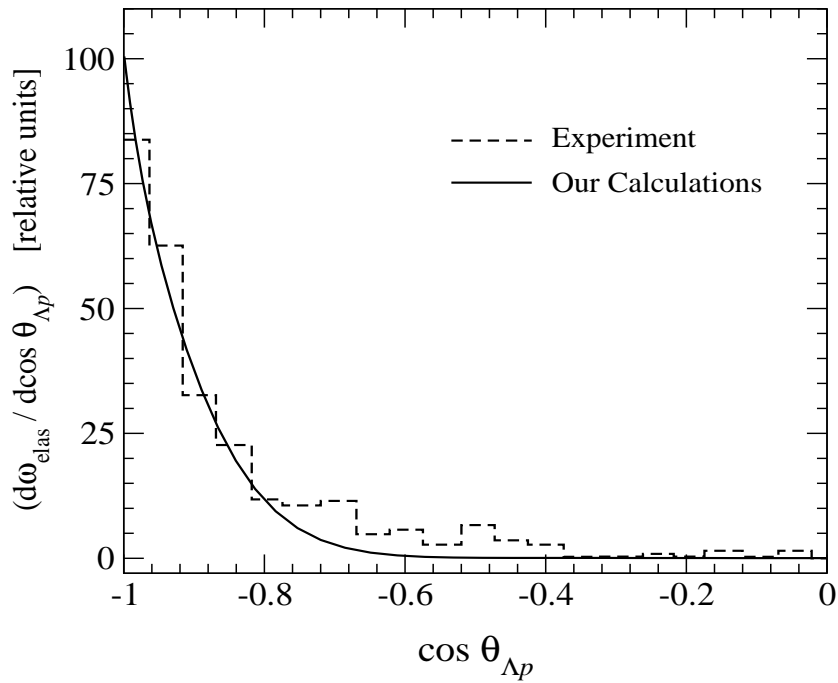}
\end{center}
\caption{Angular Correlation distribution between $\Lambda$ and 
$p$ along with the FINUDA measurements \cite{agnello}.}
\end{figure}

Measurements have also been done in the FINUDA experiment on the 
momentum distributions of the proton and $\Lambda$. The measured 
$\Lambda$ distribution from Ref. \cite{momspe} is shown in Fig. 5. 
If one believes that the observed $p$ and $\Lambda$  come from 
the two-proton absorption vertex and  do not undergo any further 
FSI except the elastic scattering, this distribution should be 
similar to the one calculated using Eq. (31). We show the calculated 
$\Lambda$ and proton momentum distributions in Fig. 5, and find that 
the peak position of the $\Lambda$ distribution  nearly agrees with 
the corresponding measured distribution. The shape of the calculated 
distributions is however found to be less broad compared to the 
observed one. Larger magnitude of measured events below 400 MeV/c, 
which makes it broader  can not be understood easily. They can not 
be attributed to the quasifree process either because the FINUDA 
spectrometer cuts off $\Lambda $'s below 300 MeV/c. Another source 
of this deficiency can be that, in our calculations we have taken 
$g_{abs}(q)$ (denoted by $C$ in Eq. (31)) as constant. We examined 
it. As we see in the phase space expression (Eq. (32)),  calculation 
of the absorption probability for each value of the $|\vec{p_\Lambda}|$ 
involves an integral over $cos\,(\theta _{p\Lambda})$, which in the 
FINUDA measurements lies between $-$1.0 and $-$0.8. This implies in 
the calculation for each $p_\Lambda$ an integral over a certain 
range of $q$ and $Q$ corresponding to this range of $p\Lambda$ angle. 
For 400, 500 and 600 MeV/c values of $p_\Lambda$ (which more or less 
covers Fig. 5 ) this range of $q$ is about 510-495, 520-505 and 
515-500 MeV/c respectively. These values, as we see are large and 
have about same range for all the $\Lambda$ momenta. Therefore our 
assumption about the constancy of $g_{abs}(q)$ in the calculations 
should not be the cause of concern. The shape of the $p_\Lambda$ 
distribution is, in fact, mainly determined by $G(Q)$ through the 
$Q$ dependence of the nuclear wave functions. The range of $Q$ for 
the above three $p_\Lambda $'s in the $-1\,\leq\, cos\,
\theta_{p\Lambda}\,\leq\,-0.8$ are 200-260, 35-220, and 160-270 MeV/c 
respectively. This range shows why the $p_\Lambda$ momentum 
distribution peaks around 500 MeV/c. It also suggests that it can 
probably be made broader by enriching the nucleon shell model wave 
functions in high momentum components. However, to reproduce the 
width of the observed momentum distribution, we believe that it will 
require a considerable modification of the nuclear wave function.

\begin{figure}[ht]
\begin{center}
\includegraphics[width=11cm,height=9cm]{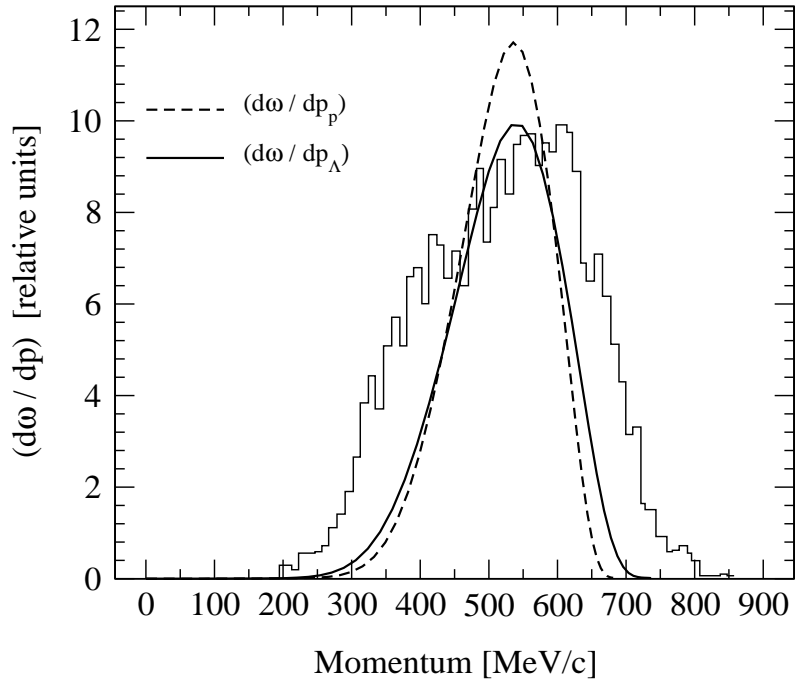}
\end{center}
\caption {$\Lambda$ and $p$ inclusive momentum distributions with 
$-1\,\leq$ cos $\theta_{p\Lambda}\,\leq\,-0.8$ along with the 
measured $\Lambda $ momentum distribution (represented by 
Histogram).}
\end{figure}

Next we calculate the  $p$-$\Lambda $ invariant mass 
distribution. The phase space for this is written as 
\begin{equation}
PS\,=\,\frac{4\pi\,\mu_{\Lambda\,p}\,\mu_{(\Lambda\,+\,p)\,B}}
{(2\pi)^5}\,|\vec{Q}|\,|\vec{q}|\,dM_{\Lambda p}\,d\,cos\,
(\theta_ {Qq}), 
\end{equation}
where $\mu_{xy}$ denotes the reduced mass of the $x,\,y$ system. 
The magnitudes of $\vec{Q}$ and $\vec{q}$ are determined for a given 
invariant mass $M_{\Lambda p}$ through 
\begin {equation}
|\vec{q}|\,=\,\sqrt{2\,\mu_{\Lambda p}\,(M_{\Lambda p}\,-\,m_p\,-\,
m_\Lambda)},
\end {equation}
and
\begin {equation}
|\vec{Q}|\,=\,\sqrt{2\,\mu_{(\Lambda\,+\,p)\,B}\,(m_{K^-}\,+\,2\,m_p
\,-\,B_{l_1\,l_2}\,-\,M_{\Lambda p})}.
\end {equation}
The angle $\theta_{Qq}$ between $\vec{Q}$ and $\vec{q}$ is constrained 
such that $-1\,\leq\,cos\,\theta_{p\Lambda}\,\leq\,-0.8$. Calculated 
invariant mass distribution along with the corresponding measured 
FINUDA distribution \cite{agnello} are shown in Fig. 6. We observe 
that, compared to the mass of $K^-\,p\,p$ in the free state (2370 MeV) 
the calculated distribution is down-shifted in mass by about 50 MeV 
due to proton binding in $^{12}$C. The measured mass distribution, 
however, is still below this by an additional 70 MeV or so. This, 
incidentally, is around the reported calculated binding energy in the 
literature of the ``$K^-\,pp$" module taking $\Lambda (1405)$ mass 
27 MeV below $K^-$-$N$ threshold.

The range of the values of $q$ and $Q$ for $2290\,\leq M_{p\Lambda}\,
\leq\,2320$ MeV, (which covers the calculated distribution) are 
520-490 and 27-320 MeV/c respectively. This says that the values of 
$q$ are large and do not vary much, hence $g(q)$ is not likely to 
influence the shape of the mass distribution. Its shape would be 
mainly determined by the nuclear wave function through $G(Q)$ and 
the factor $Q$ in the phase space.    
\begin{figure}[ht]
\begin{center}
\includegraphics[width=12cm,height=10cm]{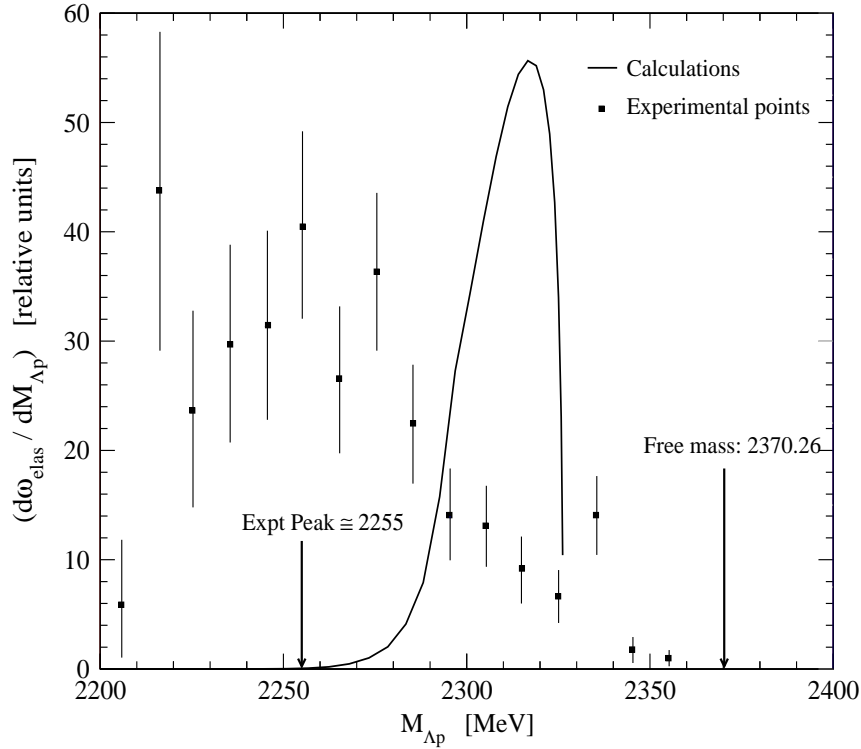}
\end{center}
\caption {Invariant mass distribution of $\Lambda $ and $p$ along 
with FINUDA measurements. $-1\,\leq$ cos $\theta_{p\Lambda}\,\leq\,
-0.8$.}
\end{figure}

If we include in the kinematics of our calculations an additional 
binding energy of 70 MeV in the initial system, then, we find that 
the peak position in the calculated mass distribution (not plotted 
here), of course, comes near to  the peak position of the measured 
distribution, but the shape of the calculated distribution turns out  
much sharper than the experimental one. The calculated $\Lambda$ 
momentum distribution with the additional binding also gets shifted 
towards lower momenta. This then spoils the agreement of the 
calculated results with the experiments shown above in Fig. 5 without 
any ``$K^-\,p\,p$" binding. The angular correlation between $\Lambda$ 
and proton, however, remains unchanged.

\subsubsection{Conclusion}
In the two-nucleon $K^-$ absorption model, with only elastic 
scattering of $p$ and $\Lambda$ included in  the final state, the 
measured $\Lambda$-$p$ angular correlation distribution is reproduced 
well, and the momentum spectrum of $\Lambda$'s to a reasonable extent 
without introducing any additional binding of the ``$K^-\,pp$" module. 
The calculated invariant $\Lambda$-$p$ mass distribution, however, 
peaks around 70 MeV higher than the measured one. Attributing this 
shift to the additional binding of the ``$K^-\,pp$" module, the 
calculations including this binding in kinematics, obviously, 
reproduce the peak position of the  measured mass distribution, but 
the shape of the distribution remains very narrow. It also spoils the 
agreement with the $\Lambda$ momentum distribution achieved without 
including any ``$K^-\,p\,p$" binding. Thus, in an overall conclusion 
it seems that the observed invariant mass distribution in the 
($K^-$, $p\,\Lambda$) reaction does not correspond to the peak 
predicted in this Section with only elastic scattering included in 
the FSI.  

\subsection{Knock-out scattering}
The knock-out scattering means that the proton or $\Lambda $, 
emanating from vertex I, encounters a hard collision with a nucleon 
in the residual nucleus and knocks it out (Fig. 2b). This collision 
alters significantly the momentum distribution of the striking 
particle. To calculate the knock-out contribution we recollect that, 
as discussed in the earlier Section, $p$ and $\Lambda $  move back 
to back at vertex I and the vertex itself is localized on the surface 
of the target nucleus. Because of this only one particle goes into the 
nucleus, the other one moves out. For such trajectories, as discussed 
earlier in Section A(2)  for distorted waves in elastic scattering, 
here too the contribution of $p$ and $\Lambda $ to KO FSI can be 
included by considering the passage of only one of the two particles, 
but through the whole nucleus. This can be reasonably correct if the 
elastic scattering parameters for $p\,N$ and $\Lambda$-$N$ systems at 
intermediate energies are not much different. Considering that it is 
so, in the following we calculate the KO contribution induced by the 
proton.    

Let $f(\vec{p_p^\prime})$ describe  the momentum distribution of the  
proton when it incidents on the vertex II, where $\vec{p_p^\prime}$ 
is the momentum of the proton at the time of leaving the vertex I. 
The altered momentum of this proton after the KO scattering is 
denoted by $\vec{p_p}$. The proton with this momentum along with the 
$\Lambda$ are detected in the experiment. The knocked-out nucleon 
from the nucleus is not seen. The probability for 
this ``inclusive" process is determined by the product of 
$f(\vec{p_p^\prime})$ and the proton induced inclusive single 
nucleon knock-out cross section as 
\begin{equation}
d\omega_{KO}\,=\,f(\vec {p_p^\prime})\,\,\frac{\sigma_{KO}(\vec{p_p},
\vec{p_p^\prime})}{\sigma_T^{pB}(p_p^\prime)},
\end{equation}
where $\sigma _T^{pB}(p_p^\prime)$ is the total proton-nucleus cross 
section at the momentum $|\vec{p_p^\prime }|$. Momentum 
$|\vec{p_p^\prime}|$ is given in terms of the variables at the first 
vertex. In terms of $\vec{p_\Lambda}$ it is given by 
\begin{equation}
|\vec{p_p^\prime}|=\mu _{p^\prime B}\,\left[\frac{-|
\vec{p_\Lambda}|\,cos\,(\theta _{\Lambda p^\prime})}{M_B} +
\left[\left(\frac{|\vec{p_\Lambda }|\,cos\,(\theta _{\Lambda 
p^\prime})}{M_B}\right)^2 -\left(\frac{2}{\mu _{pB}}\right)
\left(\frac{m_\Lambda}{\mu_{\Lambda B}}\,T_\Lambda - M\right)
\right]^{1/2}\right],                  
\end{equation}
\\
where all the notations are self explanatory. $T_x$ represents the 
kinetic energy of the particle $x$. $M=m_K+M_A-m_p-m_\Lambda-M_B$ 
. Function $ f(\vec {p_p'})$ is 
obtained from Eq. (31), by first writing the phase space $[PS]$  
 for $\vec{p_\Lambda}$ as,
\begin{equation}
[PS]\,=\,\frac{m_\Lambda\,M_B}{(2\pi)^5}\,\,\frac{|\vec{p_p^\prime}|
|\vec{p_\Lambda}|^2}{(m_\Lambda+M_B)\,|\vec{p_\Lambda}|+|
\vec {p_p^\prime}|m_\Lambda cos\,(\theta_{\Lambda p^\prime})}\,\,
d\vec {p_\Lambda},
\end{equation}    
and then substituting this phase space expression in Eq. (31), and 
identifying $f(\vec{p_p^\prime})$ with ($d\omega _{elas}\,/\,
d\vec{p_p^\prime}$). This gives  
\begin{eqnarray}
\nonumber
f(\vec{p_p^\prime})&=&\left[\frac{m_\Lambda M_B}{(2\pi)^5}
\frac{|\vec{p_p^\prime}|\,|\vec{p_\Lambda}|^2}{(m_\Lambda +M_B)
|\vec {p_\Lambda}|\,+\,|\vec {p_p^\prime}|\,m_\Lambda cos\,
(\theta _{\Lambda p'})}\,\,d\vec {p_\Lambda }\right]
\\
&&\times\,\,\left[C\,|\eta_A(T_p)|^2\,N_{l_1\,l_2}\,
\sum_L(l_1l_200/L0)^2 |g_{l_1\,l_2}^L(P_B)|^2\right]
\end{eqnarray}  

However, before we proceed further let us mention that Eq. (36) for 
the knock-out contribution, which has a factorization of the vertex 
I and II, holds under a certain approximation. More correctly, as 
shown in diagram 2(b) these two vertices should have been correlated 
in space through the proton (or lambda) propagator between them. 
This propagator would be a proton scattered wave between the two 
vertices with an outgoing boundary condition. That is, we would have 
between two vertices a factor like 
\begin{equation}
\int d\vec{p}\,\frac{f(\vec{p},\,\vec{p^\prime})}{|\vec{p^\prime }|^2
\,-\,|\vec {p}|^2\,-\,2\,m\,U\,+\,i\,\epsilon},
\end{equation}
where $f(\vec{p}, \vec {p^\prime })$ collectively represents all other 
factors. $U$ represents the proton interaction with the medium. This 
integral  has two parts, one originating from the principal value and 
another from the energy conserving $\delta$-function part of the 
propagator. Physically these two parts represent the off-shell and 
the on-shell scattering in the intermediate state. The on-shell part 
can be shown to be roughly proportional to the proton momentum, hence 
dominates at higher energies. The off-shell part dominates at lower 
energies (see for example \cite{bkj}). In our case, since the energies 
of the proton (or lambda) are in the intermediate energy range, we 
have restricted ourselves to the energy conserving on-shell 
contribution only. This, essentially is the assumption implicit in 
writing Eq. (36) with the proton being described at both vertices by 
distorted waves.    
  
\subsubsection{Inclusive Knock-out Cross section}
Proton induced single nucleon knock-out reaction at intermediate 
energies is a well studied subject experimentally as well as 
theoretically \cite{ko}. Using the notations given in diagram (2b), 
the expression for the ($p$, $p$$N$) knock-out reaction is given by 
\begin{eqnarray}
\nonumber
\frac{d\sigma}{d\vec {p_p}}&=&\frac{1}{(2\pi)^5}\,\frac{1}{v_p^\prime}
\,\,\sum_n\int d\vec{p_N}\,\,d\vec{Q_R}\,\,\delta(T_p^\prime-T_p-T_N-
T_R-E_n^*)\,\times
\\
&&\delta(\vec {p_p^\prime}-\vec{p_p}-\vec{p_N}-\vec{Q_R})\,\,
\bar{\sum}\,|\,T_n(\vec{p_p^\prime},B\rightarrow \vec{p_p},\vec {p_N},
\vec{Q_R},X_n)\,|^2,
\end{eqnarray}
where $\bar{\sum }$ denotes the average and sum over the spins in 
the initial and final states respectively. $\vec{Q_R}$ is the 
momentum of the recoiling nucleus.  $X_n$ and $E_n^*$  denote its 
intrinsic excitation and the excitation energy respectively. To 
evaluate this expression, first we integrate over 
$\vec{p_N}$ and utilize the momentum conserving delta function, 
giving
\begin{eqnarray}
\nonumber
\frac{d\sigma}{d\vec{p_p}}&=&\frac{1}{(2\pi)^5}\,\frac{1}{v_p^\prime}
\,\,\sum_n\int d\vec{Q_R}\,\,\delta(T_p^\prime-T_p-T_N-T_R-E_n^*)
\\
&&\times\,\bar{\sum}\,|\,T_n(\vec{p_p^\prime},B\rightarrow \vec{p_p},
\vec{p_N},\vec{Q_R},X_n)\,|^2,
\end{eqnarray}
with $\vec{p_N}\,=\,\vec{p_p^\prime}-\vec{p_p}-\vec{Q_R}\,=\,\vec{q}-
\vec{Q_R}$, where $\vec{q}\,=\,\vec{p_p^\prime}-\vec{p_p}$ is the 
momentum transfer from the incident proton.

The transition matrix $T_n$ in above describes the T-matrix for the 
knock-out of a nucleon from the nucleus B and leaving the residual 
nucleus in a one-hole excited state denoted by `$n$'. It is given by  
\begin{equation}
T_n(\vec{p_p^\prime},B\rightarrow \vec{p_p},\vec{p_N},X_n)\,=\,
\langle\,p,\vec{p_p};N,\vec{p_N};X_n, \vec{Q_R}\,|\,t_{p^\prime N
\rightarrow pN}(\epsilon)\,|\,p^\prime,\vec {p_p^\prime};B\,\rangle,
\end{equation}
where $B$ and $X_n$ denote the nuclear wave functions in the initial 
and final states respectively. $t_{p^\prime N\rightarrow pN}$ is the 
$N$-$N$ scattering amplitude. This amplitude is half off-shell if the 
distortion of the continuum nucleons is ignored and becomes fully 
off-shell if the distortions are included. However, at the energies 
of concern to us, the off-shell effects are known not to be 
significant. Hence, normally the $N$-$N$ t-matrix here is taken 
on-shell and the energy, $\epsilon$ at which it is evaluated is 
taken corresponding to the incident momentum $\vec{p_p^\prime}$. The 
$t_{NN}$ itself is related to the $N$-$N$ cross section in the 
centre of mass through
\begin{equation}
\frac{d\sigma}{d\Omega}\,=\,\frac{m^2/4}{(2\pi )^2}\,\,
\frac {|\vec {\kappa _f}|}{|\vec {\kappa _i}|}\,\,\bar{\Sigma }
\left|\,\langle\,\vec{\kappa_f}\,|\,t_{NN}(\epsilon)\,|\,
\vec{\kappa _i}\,\rangle\,\right|^2,
\end{equation}
where $\vec{\kappa_x}$ is the $N$-$N$ momentum in its centre of 
mass, and $m$ is the nucleon mass.

The sum over `$n$' in Eq. (42)  means the sum over the excited states
in the recoiling nucleus consistent with the momentum conservation.
Since the reaction mechanism involves only a nucleon in the nucleus 
B, `$n$' would have only a small spread. Therefore, in the energy 
delta function in Eq. (42) we can replace, to a reasonable 
approximation,  $E_n^*$ by the binding energy, $B_N$ of the 
knocked-out nucleon in B. This simplifies the summation over $n$ in 
Eq. (42). Using closure, we can then write 
\begin{equation}
\sum_n\,\,\bar{\sum }|T_n|^2\,=\,\bar{\sum }\,\,\int d\xi\,\,
\left|\,\langle\, p,\vec{p_p};N,\vec{p_N}; \vec{Q_R}\,|\,
t_{p^\prime N\rightarrow pN}(\epsilon )\,|\,p^\prime,
\vec{p_p^\prime};B\,\rangle\,\right|^2,
\end{equation}
where $\xi$ collectively denotes coordinates of all the spectator 
core nucleons in the nucleus $B$. To proceed further we now use a 
simple description of the nucleus. We write the target nucleus B 
wave function as a product of a single nucleon (`N') wave function 
in a shell with quantum numbers ``$nl$" and the core nucleus wave 
function $\Phi_c(\xi)$. With this Eq. (45) reduces to
\begin{equation}
\sum _n\,\,\bar{\sum }|T_n|^2\,=\,\bar{\sum }\,\left|\,\langle\, 
p,\vec{p_p};N,\vec{p_N}; \vec{Q_R}\,|\,t_{p^\prime N\rightarrow pN}
(\epsilon)\,|\,p^\prime,\vec{p_p^\prime};\phi_{nl}(N)\,\rangle\,
\right|^2,
\end{equation}
with 
\begin{equation}
\bar{\sum}\,=\,\frac{1}{4}\,\,\frac{S(l)}{(2l+1)}\,\,\sum_{m_l}
\,\,\sum_{nucleon-spins},
\end{equation}
where $S(l)$ is the nucleon spectroscopic factor. In the present 
simplified description of the nuclear wave function, it is equal 
to the number of nucleons in the shell ``$nl$". For the bound nucleon 
we use the momentum space representation, 
$\phi_{nl}(\vec{p_N^\prime})$, where $\vec {p_N^\prime}$ from the 
momentum conservation at the $\langle\,t_{p^\prime N\rightarrow pN}
(\epsilon)\,\rangle$ vertex and that following Eq. (42)  equals to, 
$-\,\vec {Q_R}$. With this identification Eq. (46) then factorizes as 
\begin{eqnarray}
\nonumber
\sum_n\,\,\bar{\sum}\,\,|T_n|^2&=&\left[\,\bar{\sum_{\sigma's}}\,
\left|\,\langle\,\sigma _p,\vec{p_p};\sigma _N,\vec{p_N}\,|\,
t_{pN}(\epsilon)\,|\,\sigma_p^\prime,\vec{p_p^\prime};
\sigma_N^\prime,-\vec{Q_R}\,\rangle\,\right|^2\,\right]
\\
&&\times\,\,\left[\,\frac{1}{4}\,\,\frac{S(l)}{2l+1}\,\,\sum_{m_l}
\,|\,\phi_{nlm_l}(-\,\vec{Q_R})\,|^2\,\right].
\end{eqnarray}

The single nucleon knock-out cross section  expression (Eq. (42)) 
subsequently becomes
\begin{eqnarray}
\nonumber
\frac{d\sigma}{d\vec {p_p}}&=&\frac{1}{(2\pi)^5}\,\,\frac{1}
{v_p^\prime}\,\,\int d\vec{Q_R}\,\,\delta (T_p^\prime-T_p-T_N-
T_R-B_N)
\\
\nonumber
&&\times\,\,\left[\,\bar{\sum_{\sigma's}}\,\left|\,\langle\,
\sigma_p,\vec{p_p};\sigma _N,\vec{p_N}\,|\,t_{pN}(\epsilon)\,|\,
\sigma_p^\prime,\vec{p_p^\prime};\sigma_N^\prime,-\vec{Q_R}\,
\rangle\,\right|^2\,\right]
\\
&&\times\,\,\left[\,\frac{1}{4}\,\,\frac{S(l)}{2l+1}\,\,\sum_{m_l}
\,|\,\phi_{nlm_l}(-\vec{Q_R})\,|^2\,\right].
\end{eqnarray}

To obtain the expression for the inclusive cross section we still 
need to integrate this expression over $\vec {Q_R}$. Following 
Ref. \cite{wolff} we use the energy delta function to remove angle 
integration, and with some algebraic manipulations obtain
\begin{eqnarray}
\nonumber
&&\int d\vec{Q_R}\,\,\delta(T_p^\prime-T_p-T_N-T_R-B_N)\,\left[\,
\frac{S(l)}{2l+1}\,\,\sum_{ml}\,|\,\phi_{nlm_l}(-\vec {Q_R})\,|^2\,
\right]
\\
&&\hspace{4cm}=\,S(l)\,\,\frac{m}{2q}\,\,
\int_{Q_R^{min}}^\infty Q_R\,\,dQ_R\,\,|\,\phi _{nl}(Q_R)\,|^2.
\end{eqnarray} 
Substituting this in  Eq. (49) above, and also writing 
$\displaystyle\bar{\sum_{\sigma's}}\,\left|\,\langle\,t_{pN}\,
\rangle\,\right|^2$ in terms of the elementary $N$-$N$ differential 
cross section (Eq. (44)) we get 
\begin{equation}
\frac{d\sigma}{d\vec{p_p}}\,=\,\left[\,\frac{2}{qp_p^\prime}\,\right]
\,\,\left[\,\frac{d\sigma}{d\Omega}(\epsilon)\,\right]_{cm}^{pN}\,\,
\left[\,S(l)\,\int_{Q^{min}_R}^\infty Q_R\,\,dQ_R\,\,\left|\,
\frac{1}{(2\pi)^{3/2}}\,\,\phi _{nl}(Q_R)\,\right|^2\,\right].
\end{equation}
Here $Q^{min}_R$ is that minimum momentum which a nucleon of 
binding energy $B_N$ must have in the nucleus for the scattered 
proton to be observed at a scattering angle $\theta$ with a 
momentum $|\vec{p_p}|$. Its value  is given by 
\begin{equation}
Q^{min}_R\,=\,\frac{[\,p_p^2\,-\,p_p^\prime\,p_p\,cos\,(\theta)\,+\,
m\,B_N\,]}{[\,p_{p^\prime}^2\,-\,p_p^\prime\,p_p\,cos\,(\theta)\,+\,
m\,B_N\,]^{1/2}}\,\,.
\end{equation}

Eq. (51) for the knock-out cross section assumes that the incoming 
proton and the outgoing nucleons do not suffer any additional 
scattering except the hard knock-out collision. This additional 
scattering, however, can be incorporated in the formalism by 
replacing the plane wave description of the continuum nucleons by 
the ``distorted waves", which would be solutions of the wave 
equation with appropriate optical potentials in it. Several studies 
of the distortion effect in the knock-out reaction have been carried 
out in the literature and it has been found that in the energy range 
of nucleons of interest to us here, the effect of distortion is 
mainly absorptive. The dispersive effect is very small. The 
absorption factor for $^{12}$C ($p$, $2p$) $^{11}B$ reaction at 
160 MeV beam energy, for example, in Ref. \cite{kwall} is found to 
be around 0.5.

Finally, before closing this Section  we determine the extent of 
accuracy to which the expression in Eq. (51) describes the measured 
inclusive proton induced single nucleon knock-out cross section. We 
calculate the inclusive cross sections at 160 MeV beam energy for 
$^{12}$C and compare them with the measured ($p$,$p^\prime $) ones at 
the same beam energy \cite{wall}. The calculated results are summed 
over the knocked-out nucleon (including both neutron and proton) 
from $1s$ and $1p$ shells. The single nucleon binding energies for 
them are taken from Ref. \cite{be}. The nucleon-nucleon differential 
cross section in the centre-of-mass required in the calculations 
are taken from the analytic form given in Ref. \cite{riley}, i.e.
\begin{equation}
\frac{d^2\sigma}{d\Omega dT}\,=\,\left[\,1.9\,+\,\frac{230}{T}\,+\,
\frac{4850}{T^2}\,\right]\,(\,1\,+\,0.1\,cos^2\,\theta\,),
\end{equation}
where the cross section is in mb and energy (kinetic), T in MeV. 
This form is in good agreement with the energy dependence 
of the observed cross sections for the range 20 $\le$ T $\le$ 
200 MeV.

The bound nucleon wave function is described by the  oscillator 
potential form with the length parameter, b = 1.67 fm. The 
spectroscopic factor, $S(l)$  is taken equal to the number of 
nucleons (neutrons+protons) in the orbital $nl$.
\begin{figure}[ht]
\begin{center}
\includegraphics[width=11cm,height=9cm]{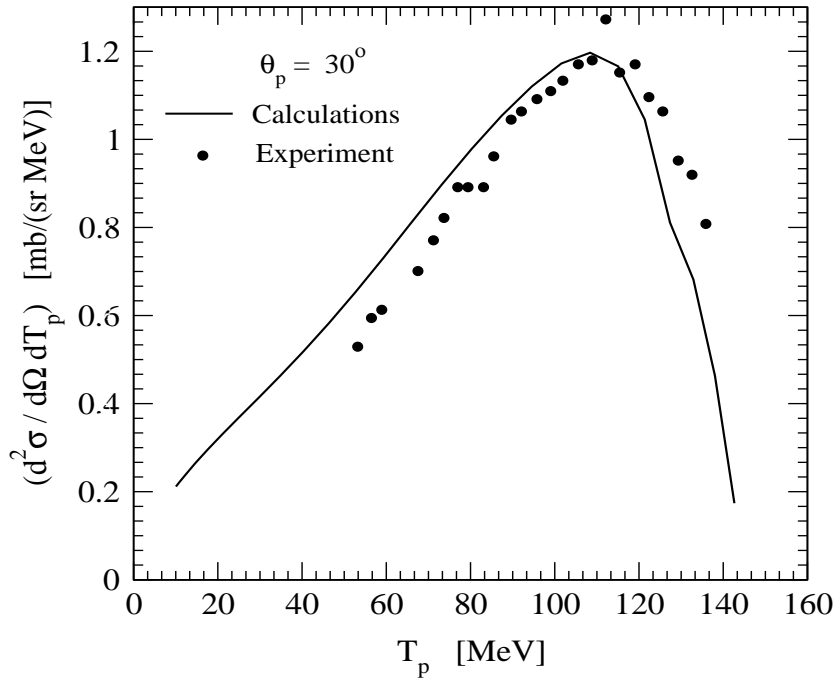}
\end{center}
\caption {Calculated energy spectrum of a 160 MeV incident proton 
on $^{12}$C after a single nucleon knock-out scattering from the 
nucleus along with the experimental points at the same energy 
\cite{wall}. }
\end{figure}

With these inputs and the distortion factor equal to 0.5 (as 
discussed above), the calculated energy spectrum for the proton 
going at 30$^\circ$, for example, in the lab. frame  along with the 
experimental cross sections is shown in Fig. 7. As we see, the 
agreement between them, both in shape and magnitude, is very good. 
This validates the accuracy of Eq. (51) for the description of the 
inclusive knock-out channel and gives confidence for its use in 
calculations  of the ($K^-$, $p\,\Lambda$) reaction.

\subsubsection{Results}
The final expression for calculating the knock-out contribution to 
the $K^-$ absorption probability is obtained  by substituting 
Eq. (51) for $\sigma_{KO}(\vec{p_p},\vec{p_p^\prime})$ in Eq. (36). 
We get  
\begin{eqnarray}
\nonumber
d\omega _{KO}&=&[d\vec{p_\Lambda}\,\,d\vec{p_p}]\,\,
f(\vec {p_p^\prime})\,\times\,\frac{1}{\sigma _T^{pB}(T_{p^\prime})}
\,\,\left[\,\frac {2}{qp_p^\prime}\,\right]\,\,\left[\,\frac{d\sigma}
{d\Omega}(\epsilon)\,\right]_{cm}^{pN}\times 
\\
&&\left[\,S(l)\,\int_{Q^{min}_R}^\infty Q_R\,\,dQ_R\,\left|\,
\frac{1}{(2\pi )^{3/2}}\,\,\phi _{nl}(Q_R)\,\right|^2\,\right],
\end{eqnarray}              
where $f(\vec {p_p'})$ is given by Eq. (39). As in the case of 
``elastic scattering" we present here calculated results for the 
target nucleus $^{12}$C. The results are the probabilities summed 
over two proton hole states in various pairs of ($n_1\,l_1$; 
$n_2\,l_2$) shell model orbitals at vertex I and, for each of these 
pairs, summed over various  one nucleon $nl$ orbitals in nucleus $B$ 
at the knock-out vertex II. Since in the final state after knock-out 
we do not detect the knocked out nucleon, we use the spectroscopic 
factor, $S(l)$ summed  over both, the  neutrons and protons. The 
radial part of the bound state wave functions, as discussed in the 
last Section, is taken for the oscillator potential. The 
differential scattering cross section for $p$-$N$ is described by 
Eq. (53) with energy  taken corresponding to momentum 
$\vec {p_p^\prime}$ of the proton incident at the vertex II 
(Fig. 2b).  
Required $\sigma _T$ for proton on nucleus $B$ at energies 
corresponding to different proton momentum, $|\vec{p_p^\prime}|$ is 
taken from Ref. \cite{devries}. This cross section, of course, does 
not vary significantly over the $p^\prime$ energy range of interest 
here.

The physical effect of knock-out scattering at vertex II is to 
reduce the energy of the proton $p^\prime$ and deflect it from its 
direction of incidence. The amount of these changes, as can be 
seen from the experimental results on inclusive ($p$, $p^\prime$) 
reaction at 160 MeV in Ref. \cite{wall}, are about 30 MeV and above 
for the energy reduction and about 30$^\circ$ and above for the 
deflection. Immediate consequence of these numbers would be that 
the angular correlation between $p$ and $\Lambda$ shown in Fig. 4, 
coming from vertex I, will be widened significantly and the energy 
spectrum of $p$ and $\Lambda$, shown in Fig. 5, will be shifted 
towards lower energies. Both these effects will, therefore, spoil 
the agreement shown in these figures (Figs. 4-5) between the 
calculated results from vertex I and the corresponding FINUDA 
measurements.
\begin{figure}[ht]
\begin{center}
\includegraphics[width=11cm,height=9cm]{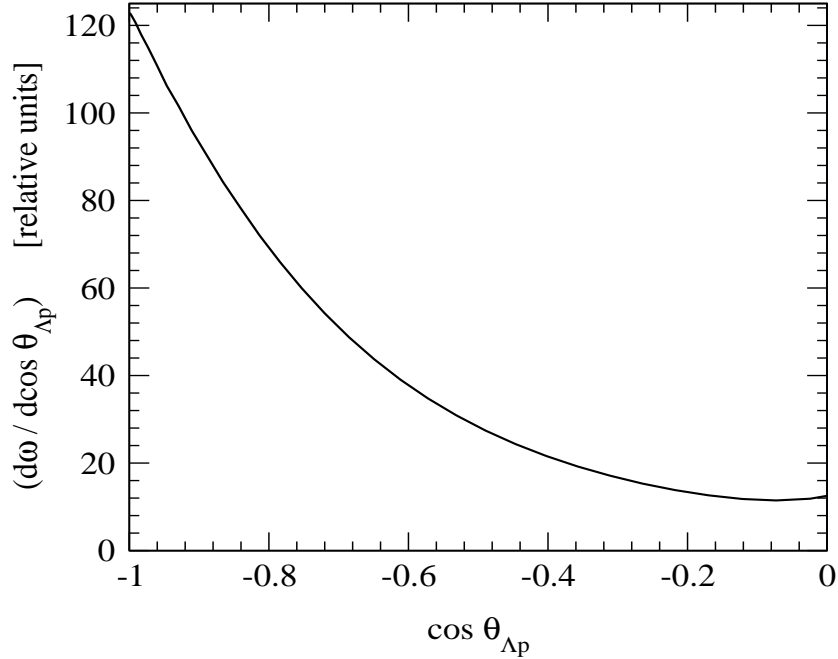}
\end{center}
\caption {Angular distribution of an incident proton after a 
single nucleon knock-out scattering on vertex II.}
\end{figure}
To get a quantitative  idea about the extent of deflection the 
vertex II may introduce in the proton, $p^\prime$, in Fig. 8 we show 
its angular distribution relative to the $\Lambda $ motion after 
being scattered from the vertex II. Initially the $p\prime $ is taken 
to move at 180$^\circ$ w.r.t. $\Lambda$ with energy as fixed at the 
vertex I, including the spread  due to Fermi motion.  
Without the knock-out scattering, this distribution  will be just a 
point at cos($\theta_{p\Lambda}$)=$-$1 in this figure. Due to 
scattering this point gets a significant spread, as we see in this 
figure. Calculated results use Eq. (54) and are integrated over the 
energy spread of $p^\prime$.      
\begin{figure}[ht]
\begin{center}
\includegraphics[width=11cm,height=10cm]{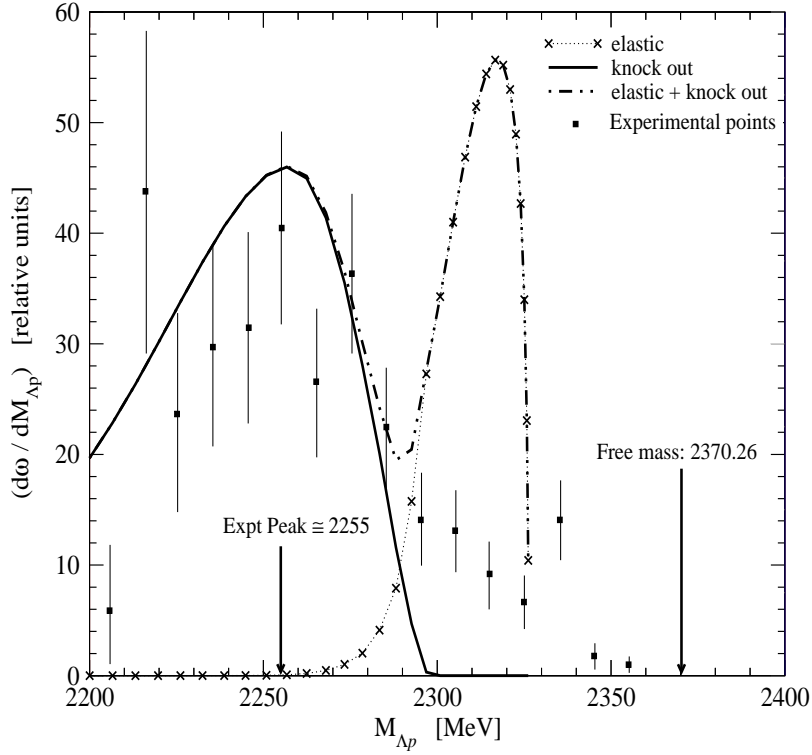}
\end{center}
\caption {Invariant $p\Lambda$ mass distribution from both the 
vertices along with the experimental points \cite{agnello}.}
\end{figure}
  
The consequence of above is expected to be a significant modification 
in the invariant mass distribution of the $p$ and $\Lambda $. However,   
before we show these results we  may mention that, because of various 
sums, integrals and constrain checks on kinematic variables, the 
calculations are very involved and tedious. Therefore, to keep the 
calculations somewhat simpler and  physically transparent we have 
put some constraints in the calculations without, of course, losing 
any essence of the physics of the results. We have seen earlier  
in the calculations on vertex I that the $p$ and $\Lambda $ from it 
emerge mainly back-to-back with a very small cone angle. We have, 
therefore, done the vertex II calculations with $\theta_{p\Lambda}$
 = 180$^{\circ}$ only. The energy variation of the $p$ and $\Lambda$ 
due to Fermi motion of the absorbing protons in the nucleus A, 
however, has been kept intact. With this, in Fig. 9 we show the 
calculated invariant mass distribution of $p$-$\Lambda$ after 
knock-out scattering along with the FINUDA measurements 
\cite{agnello}. It is extraordinary to see that the calculated mass 
distribution totally agrees with the measured ones. They agree in 
mass shift as well as in the shape of the distribution. This is the 
same observation as reported in Ref. \cite{magas}.

However, to get the complete $p$-$\Lambda $ invariant mass 
distribution we need to add to above the contribution from the 
``elastic scattering" FSI corresponding to vertex I (Fig. 2a). 
Therefore, in Fig. 9 we also show the $p$-$\Lambda$ invariant mass 
distribution due to vertex I, and the sum of the ``elastic" and 
``knock-out" contributions. The summed distribution obviously has 
two peaks, one corresponding to ``elastic scattering" FSI and another 
to the ``single nucleon knock-out" FSI. This is similar to the 
inclusive inelastic spectrum normally seen experimentally in 
($p$, $p^\prime$) or ($e$, $e^\prime$) scattering on any nucleus at 
``intermediate energies" \cite{wall}. Unusual thing following $K^-$ 
absorption seen here is that the peak corresponding to ``elastic 
scattering" seems to be missing in the FINUDA data. This is 
intriguing.

\subsubsection{Conclusions} 
We summarize our observations for the Vertex II as,
\begin{enumerate}
\item The calculated invariant $p$-$\Lambda$ mass distribution 
totally reproduces the experimentally observed distribution. This 
is in line with the finding in Ref. \cite{magas}.
\item Though we have not shown the full calculations for the 
inclusive energy spectra for the proton or lambda and angular 
correlation between them, we believe that due to large deflection 
and energy shift by the single-nucleon knock-out scattering the 
agreement seen in Figs. 4-5 between the observed and calculated 
angular correlation between the $p$ and $\Lambda $ and the inclusive 
$p$ and $\Lambda $ energy spectra using only the vertex I will be 
spoiled considerably. We have not calculated these spectra fully 
because, due to several integrations over various kinematic 
variables, they are very long and involved. 
\end{enumerate}  

\section{Summary and Final Conclusions}
We have calculated the inclusive differential absorption 
probability for $K^-$ at rest in $^{12}$C nucleus for the 
$p\,\Lambda$ exit channel. The $K^-$ is assumed to be absorbed on 
a pair of protons in the nucleus. The final state interaction in the 
reaction includes the elastic scattering of $p$ and $\Lambda $ in 
the final state and the single nucleon knock-out from the recoiling 
nucleus. The calculated invariant $p$-$\Lambda$ mass 
distribution shows two peaks, one due to elastic scattering and 
another due to knock-out channel. The latter peak overlaps in 
position and width with the peak observed in the FINUDA measurements. 
The peak corresponding to elastic scattering is not seen in 
the experiments.

Measured angular correlation between $p$ and $\Lambda $ and their 
inclusive energy distribution agree with the corresponding 
calculated results including only the elastic scattering in the final 
state. Inclusion of the knock-out channel is likely to spoil this 
distribution.

Thus, finally, we may conclude that, seen in isolation, the
experimentally observed  shift in the invariant $p$-$\Lambda$ mass 
distribution could be interpreted as due to single nucleon knock-out 
final state interaction. But, if we include other results, like the 
absence of elastic scattering peak in experiments, full agreement of 
the calculated $p$-$\Lambda $ angular correlation and their inclusive 
spectra using only vertex I with the corresponding measurements, 
the situation becomes quite a bit confusing. 

As a final comment in the present work on the ``$K^-\,p\,p$"
cluster interpretation of the observed downshift of about 100 MeV
in the FINUDA measurements of the $\Lambda\,p$ invariant mass compared
to its free value, the knock-out reaction in the final state seems
to be a definitive alternative for this shift. Only discomfiture in
this conclusion comes from the absence of the ``elastic scattering"
peak (Fig. 9) in the observed invariant mass distribution in the
FINUDA measurements. This absence can not be attributed to the cut
off of the $\Lambda$ hyperons below 300 MeV/c momentum in the
FINUDA spectrometer as these momenta in the ``elastic scattering"
peak are above 400 MeV/c.

Availability of absolute measurements may help to understand the 
($K^-$, $p\,\Lambda$) reaction better.

\section{Acknowledgements}
The authors are extremely grateful to Dr. Neelima Kelkar and 
Dr. Kanchan Khemchandani for many useful suggestions and their 
comments on the presentation of the work. This work was initiated 
sometime back when BKJ visited the Institute of Particle and Nuclear 
Studies, KEK, Japan. He thanks Prof. Shunzo Kumano and other members 
of the Institute for many useful academic discussions and their 
hospitality during the stay. The work was done under the financial 
grant from the Department of Science and Technology, Govt. of India.


\begin{thebibliography}{99}
\bibitem{agnello}
M. Agnello {\it et al.}, Phys. Rev. Lett. {\bf 94}, 212303 (2005).

\bibitem{panic08}
H. Fujioka {\it et al.}, Nucl. Phys. {\bf A827}, 303c (2009).

\bibitem{disto}
T. Yamazaki {\it et al.}, Proceedings of the International 
Conference on Exotic Atoms and Related Topics and International 
Conference on Low Energy Antiproton Physics (EXA/LEAP 2008), Vienna, 
Austria, 2008;  arXiv:0810.5182v1.

\bibitem{weise1}
N. Kaiser, P. B. Siegel, and W. Weise, Nucl. Phys. {\bf A594}, 
325 (1995).

\bibitem{oolk}
B. Krippa, Phys. Rev. C {\bf 58}, 1333 (1998);
E. Oset and A. Ramos, Nucl. Phys. {\bf A635}, 99 (1998); 
J. A. Oller and U. G. Mei$\ss$ner, Phys. Lett. {\bf B500}, 263 
(2001); M. F. M. Lutz and E. E. Kolomeitsev, Nucl. Phys. {\bf A700}, 
193 (2002).

\bibitem{pdg}
C. Amsler {\it et al.} (Particle Data Group), Phys. Lett. {\bf B667},  
1 (2008) (http://pdg.lbl.gov).

\bibitem{weise2}
T. Hyodo and W. Weise, Phys. Rev C {\bf 77}, 035204 (2008).

\bibitem{zychor}
I. Zychor {\it et al.}, Phys. Lett. {\bf B660}, 167 (2008).

\bibitem{yama1}
Y. Akaishi and T. Yamazaki, Phys. Rev. C {\bf 65}, 044005 (2002).

\bibitem{dote}
A. Dote, T. Hyodo, and W. Weise, Nucl. Phys. {\bf A804}, 197 
(2008); arXiv:0802.0238.

\bibitem{shev}
N. V. Shevchenko, A. Gal, and J. Mares, Phys. Rev. Lett. {\bf 98}, 
082301 (2007); J. Revai, Phys. Rev. C {\bf 76}, 044004 (2007).

\bibitem{magas}
V. K. Magas, E. Oset, A. Ramos, and H. Toki, Phys. Rev. C {\bf 74}, 
025206 (2006).

\bibitem{wall}
N. S. Wall and P. R. Roos, Phys. Rev. C {\bf 150}, 811 (1966).

\bibitem{brown}
G. E. Brown, {\it Unified Theory of Nuclear Models and Forces} 
(North-Holland Pub., 1967), p. 201.

\bibitem{jastrow}
R. Jastrow, Phys. Rev. {\bf 98}, 1479 (1955).

\bibitem{jackson}
Daphne F. Jackson, {\it Nuclear Reactions} (Methuen $\&$ Co. Ltd., 
London, 1970), Ch. 3.10.

\bibitem{fermi}
A. G. Sitenko, {\it Theory of Nuclear Reactions} (World Scientific 
Publishing, 1990), p. 550-555.

\bibitem{den}
L. R. Suelzle, M. R. Yearian, and Hall Crannell, Phys. Rev. 
{\bf 162}, 992 (1967); B. K. Jain, Phys. Rev. C {\bf 27}, 794 (1983).

\bibitem{be}
H. Tyren, S. Kullander, O. Sundbag, R. Ramchandran, P. Isaccson, and 
T. Berggren, Nucl. Phys. {\bf 79}, 321 (1966); 
W. D. Simpson {\it et al.}, Nucl. Phys. {\bf A140}, 201 (1970). 

\bibitem{momspe}
H. Fujioka, Graduate Study Report, Nagae Laboratory, Department of 
Physics, University of Tokyo, 2005, p. 130 (unpublished).

\bibitem{ko}
G. Jacob and Th. A. J. Maris, Nucl. Phys. {\bf 31}, 139 (1962); 
Rev. Mod. Phys. {\bf 38}, 121 (1966); T. Berggren and H. Tyren, 
Ann. Rev. Nucl. Sci. {\bf 16}, 153 (1966); Daphne F. Jackson, 
Adv. Nucl. Phys. {\bf 4}, 1 (1971); R. Shanta and B. K. Jain, 
Nucl. Phys. {\bf B175}, 417 (1971); R.E. Chrien {\it et al.}, 
Phys. Rev. C {\bf 21}, 1014 (1980); J. S. O'Connell {\it et al.}, 
{\it ibid} {\bf 35}, 1063 (1987).

\bibitem {wolff}
Peter A. Wolff, Phys. Rev. {\bf 87}, 434 (1952).

\bibitem{kwall}
F. R. Kroll and N. S. Wall, Phys. Rev. C {\bf 1}, 138 (1970).

\bibitem{riley}
K. F. Riley, Nucl. Phys. {\bf 13}, 407 (1959).

\bibitem{devries}
S. Barshay, C. B. Dover, and J. P. Vary, Phys. Rev. C {\bf 11}, 
360 (1975); R.M. de Vries and J. C. Peng, Phys. Rev. C {\bf 22}, 
1055 (1980); P. Renberg {\it et al.}, Nucl. Phys. {\bf A183}, 
81 (1972).

\bibitem{back}
G. Backenstoss {\it et al.}, Nucl. Phys. {\bf 73}, 189 (1974).

\bibitem{bkj}
N. J. Upadhyay, K. P. Khemchandani, B. K. Jain, and N. G. Kelkar, 
Phys. Rev. C {\bf 75}, 054002 (2007); N. J. Upadhyay, N. G. Kelkar, 
and B. K. Jain, Nucl. Phys. {\bf A824}, 17 (2009).

\bibitem{rook}
K. Aslam and J. R. Rook, Nucl. Phys. {\bf B20}, 159 (1970). 
\end{thebibliography}
\end{document}